\documentclass[submission, Phys]{SciPost}
\hypersetup{
    colorlinks,
    linkcolor={red!50!black},
    citecolor={blue!50!black},
    urlcolor={blue!80!black}
}

\usepackage[bitstream-charter]{mathdesign}
\DeclareSymbolFont{usualmathcal}{OMS}{cmsy}{m}{n}
\DeclareSymbolFontAlphabet{\mathcal}{usualmathcal}
\usepackage[T1]{fontenc} % if needed
\usepackage[textsize=footnotesize,textwidth=2.5cm]{todo notes}
\usepackage{cleveref}
\usepackage{caption}
\usepackage{subcaption}
\usepackage{physics}
\usepackage{tikz}
\usepackage{tikz-cd}
\usepackage{braket}
\usepackage{bm}

\newcommand{\sqbra}[1]{\left[#1\right]}

\def\n3{\sqrt{3}}

\def\D{\mathrm{d}}
\def\is{\mathrm{Is}}
\renewcommand{\vec}[1]{\mathbf{#1}}
\usepackage{simplewick}
\usepackage{color}
\usepackage{framed}
\usepackage{titlesec}
\allowdisplaybreaks[4]
\usepackage{colortbl}
\definecolor{LouisBlue}{RGB}{55, 114, 202}
\definecolor{LouisOrange}{RGB}{214, 110, 42}
\usetikzlibrary{decorations.markings}
\tikzset{middlearrow/.style={
        decoration={markings,
            mark= at position #1 with {\arrow{latex}} ,
        },
        postaction={decorate}
    }
}
\usepackage{hyperref}
\hypersetup{
    colorlinks=true,
    linkcolor=LouisBlue,
    citecolor=LouisBlue,
    urlcolor=LouisBlue,
} 

\usepackage{array}    
\usepackage{amsmath}

\begin{document}
 \hfill USTC-ICTS/PCFT-23-36\\

\begin{center}{\Large \textbf{Cutoff brane vs the Karch-Randall brane: the fluctuating case}}\end{center}

% TODO: write the author list here. Use initials + surname format.
% Separate subsequent authors by a comma, omit comma at the end of the list.
% Mark the corresponding author with a superscript *.
\begin{center}
Jiong Lin\textsuperscript{1,2,3 $\ast$}, Yizhou Lu\textsuperscript{4 $\ast$} and Qiang Wen\textsuperscript{5 $\dagger$}

\end{center}

\begin{center}
{\bf 1} School of Physics and Information Engineering, Guangdong University of Education, Guangzhou 510303, China\\
{\bf 2} Interdisciplinary Center for Theoretical Study, University of Science and Technology of China, Hefei, Anhui 230026, China\\
{\bf 3} Peng Huanwu Center for Fundamental Theory, Hefei, Anhui 230026, China\\
{\bf 4}  Department of Physics, Southern University of Science and Technology, Shenzhen 518055, China\\
{\bf 5 } Shing-Tung Yau Center and  School of Physics, Southeast University, Nanjing 210096, China

$\ast$ These authors contribute equally
\\
$\dagger$ Corresponding to: { wenqiang@seu.edu.cn}

%{ debarshi@iitk.ac.in,~wenqiang@seu.edu.cn,~sjzhou@whu.edu.cn}
\end{center}

\section*{Abstract}
Recently, certain holographic Weyl transformed CFT$_2$ is proposed to capture the main features of the AdS$_3$/BCFT$_2$ correspondence \cite{Basu:2022crn,Basu:2023wmv}. In this paper, by adapting the Weyl transformation, we simulate a generalized AdS/BCFT set-up where the fluctuation of the Karch-Randall (KR) brane is considered. In the gravity dual of the Weyl transformed CFT, the so-called cutoff brane induced by the Weyl transformation plays the same role as the KR brane. Unlike the non-fluctuating configuration, in the $2d$ effective theory the additional twist operator is inserted at a different place, compared with the one inserted on the brane. Though this is well-understood in the Weyl transformed CFT set-up, it is confusing in the AdS/BCFT set-up where the effective theory is supposed to locate on the brane. This confusion indicates that the KR brane may be emergent from the boundary CFT$_2$ via the Weyl transformations. 

We also calculate the balanced partial entanglement (BPE) in the fluctuating brane configurations and find it coincide with the entanglement wedge cross-section (EWCS). This is a non-trivial test for the correspondence between the BPE and the EWCS, and a non-trivial consistency check for the Weyl transformed CFT set-up.
%----------------------------------------------------------------------------------------
% Literature review
%----------------------------------------------------------------------------------------
\newpage
\setcounter{tocdepth}{2}
{
\hypersetup{linkcolor=LouisBlue}
\vspace{10pt}
\noindent\rule{\textwidth}{1pt}
\tableofcontents
\noindent\rule{\textwidth}{1pt}
\vspace{10pt}
}

\section{Introduction}
In AdS/CFT correspondence \cite{Maldacena:1997re}, the entanglement entropy $S_{A}$ for a region $A$ on the boundary CFT is proposed to be captured by the area of the minimal surface $\mathcal E_A$ homologous to $A$ in the AdS bulk. This is the famous Ryu-Takayanagi (RT) formula \cite{Ryu:2006bv,Hubeny:2007xt}
\begin{equation}
    S_A=\frac{\text{Area}(\mathcal{E}_A)}{4G}.
\end{equation}
Taking account of the quantum corrections from bulk fields, the RT formula was then refined to the quantum extremal surface (QES) formula \cite{Lewkowycz:2013nqa,Engelhardt:2014gca}.
The QES formula can be applied to compute the entanglement entropy of the Hawking radiation after Page time \cite{Almheiri:2019psf,Penington:2019npb}, and remarkably a region in the black hole interior could be regarded as part of the Hawking radiation. Such a region is called the entanglement island, and QES formula in this context was refined to be the island formula \cite{Almheiri:2019hni,Penington:2019kki,Almheiri:2019qdq}\footnote{See \cite{Marolf:2020xie,Basu:2023wmv,Lu:2022cgq,Lu:2021gmv,Basu:2022crn,KumarBasak:2020ams,Yu:2023whl,Miao:2022mdx,Li:2023fly,Geng:2020qvw,Deng:2020ent,An:2023dmo,Gan:2022jay,Hartman:2020khs,Hashimoto:2020cas,Ling:2020laa,Akal:2020twv,Wang:2021woy,Guo:2023fly,Chang:2023gkt,Afrasiar:2022fid,Geng:2023iqd,Deng:2023pjs} for other recent processes in island.},
\begin{align}\label{island_EE}
    S(\mathcal{R})=\min \text{ext}_{\text{Is} (\mathcal{R})}\sqbra{\frac{\text{Area}(\partial\text{Is} (\mathcal{R}))}{4G_N}+S_{\rm bulk}(\mathcal{R}\cup \text{Is} (\mathcal{R}))
    }\,,
\end{align}
Where $\mathcal{R}$ represent the Hawking radiatin and $\is (\mathcal{R})$ is the corresponding entanglement island. In \cite{Penington:2019kki,Almheiri:2019qdq}, the island formula was derived from the gravitational path integral, where a new geometric saddle called the replica wormhole dominates the path integral after the Page time.

The island formula can be realized in the 2$d$ effective description of AdS$_3$/BCFT$_2$ set-ups \cite{Deng:2020ent,Suzuki:2022xwv} via Karch-Randall braneworld holography \cite{Randall:1999ee,Randall:1999vf,Karch:2000ct}. 
By braneworld holography, one gets the induced gravity on the brane.
When the brane locates at the constant $\rho$ slice where $\rho$ is the hyperbolic angle, 
the induced gravity is topological \cite{Deng:2020ent,Suzuki:2022xwv}.
If we let the brane fluctuate in the bulk and keep the first order fluctuation $\rho=\rho_0+\delta\rho$, then the localized gravity on the brane coupled to the scalar degree describing the fluctuation is precisely the JT gravity \cite{Geng:2022slq,Geng:2022tfc,Deng:2022yll}.
The induced gravity on the brane connected with a non-gravitational CFT bath with transparent boundary conditions, provides a 2$d$ effective description of AdS/BCFT. 
Then by applying island formula on this $2d$ boundary, 
the entanglement entropy is exactly the same as the one from $3d$-bulk RT-like formula.

Very recently, it was proposed in \cite{Basu:2022crn} that the island formula can be understood from a pure quantum information perspective based on certain non-trivial constraints on the Hilbert space. More explicitly the constraints that induce  entanglement islands should be understood as projecting out certain states in the Hilbert space, such that for all the states remaining in the reduced Hilbert space, there exists a mapping from the state of the subset $\mathcal{R}$ to the state of the subset $I$, which we call a coding relation. In other words, with the Hilbert space properly reduced, given the state of $\mathcal{R}$ one can determine the state of $I$ through the coding relation. It is important the stress that, the constraints are not just imposed on one state of the system, they are imposed to the whole Hilbert space, such that for any state reduced Hilbert space, it should be confined in this reduced Hilbert spaces under any operations or evolution. Under such constraints, spacelike separated degrees of freedom on a Cauchy slice could depend on each other, which is a relation goes beyond causality. {  In non-gravitational quantum systems, such constraints are external, while in gravitational systems the Hilbert space reduces and the coding relation emerges when space-time wormhole structure emerges in the replica manifold, hence the constraints are an intrinsic property of gravity \cite{Penington:2019kki,Almheiri:2019qdq}.} 

We call a system \emph{self-encoded} if the Hilbert space is properly reduced following the above described constraints. Consider the simple case with one pair of $\mathcal{R}$ and $I$ region in the system, and the state of $I$ is totally encoded in the state of $\mathcal{R}$, the coding relation can be simply summarized by the following equation,
\begin{align}\label{core}
						\ket{i}_{I}=f(\ket{j}_{\mathcal{R}})\,.
 \end{align} 
In the path integral formalism, when we compute the elements of the reduced density matrix of $\mathcal{R}$, we cut the region $\mathcal{R}$ open and set different boundary conditions on the open edges. In this case, due to the self-encoding property, we should simultaneously cut $I$ open and set boundary conditions following the coding relation \eqref{core}. One of the important consequence is that, we can only trace the degrees of freedom in the complement of $\mathcal{R}\cup I$. The other is that, additional twist operators are inserted at the boundary of $I$. Eventually we arrive at the following formula for the entanglement entropy of $\mathcal{R}$ \cite{Basu:2022crn},
\begin{align}\label{islandf2}
S({\mathcal{R}})&=\tilde{S}({\mathcal{R}\cup I})
\cr
&=\frac{\mathrm{Area} (\partial I)}{4G_N}+\tilde{S}_{\text{bulk}}(\mathcal{R}\cup I)\,.
\end{align}
In the above equation $\tilde{S}(A)$ represents the von Neumann entropy of the reduced density matrix calculated by tracing out the degrees of freedom in the complement of $A$.
Like the configurations where the island formula \eqref{island_EE} applies, in the second equality we have assumed that the region $I$ is settled in gravitational background. 
Hence the von Neumann entropy consists of two parts, the gravitational contribution $\frac{\mathrm{Area} (\partial I)}{4G}$ and the bulk entanglement entropy $\tilde{S}_{\text{bulk}}(\mathcal{R}\cup I)$ \cite{Lewkowycz:2013nqa,Faulkner:2013ana}. As was pointed out in \cite{Basu:2022crn}, the formula \eqref{islandf2} coincide with the island formula \eqref{island_EE}, with the only difference that \eqref{island_EE} is an optimization while \eqref{islandf2} is fixed by the coding relation \eqref{core}. The difference is understandable since the coding relation that result in the formula \eqref{island_EE} should be much more complicated than the simple relation \eqref{core}. Later we also call $I$ the entanglement island of $\mathcal{R}$ and denote it as Is$(\mathcal{R})$, and refer to the formulas \eqref{island_EE} and \eqref{islandf2} the island formula $I$ and island formula $II$ respectively. 

{ In \cite{Basu:2022crn} it was discussed that, gravitational theories are self-encoded due to the emergence of spacetime wormholes. Also the Hilbert space is vastly reduced\footnote{This reduction means to compare with the Hilbert space of a quantum field theory in curved geometry background with gravitational fluctuations when the quantum effects of gravity are not prominent.} and the cutoff scale becomes finite.} Inspired by the self-encoding property and the island formula $II$, the authors of \cite{Basu:2022crn} proposed that, a holographic CFT$_2$ under certain Weyl transformation which introduces finite cutoff by erasing the UV physics of the theory in certain regions, could be a perfect simulation for the Hilbert space reduction of the gravity. More explicitly, the so-called cutoff brane in the Weyl transformed CFT$_2$ set-up plays the role of the KR (or end of world) brane in the AdS$_3$/BCFT$_2$ set-up \cite{Takayanagi:2011zk}, and the entanglement structure of the Weyl transformed CFT$_2$ perfectly reproduces the island formula $I$ in the AdS/BCFT set-up.

For a given region $A$, if the KR brane in the bulk fluctuates, the RT surface $\mathcal{E}_{A}$, entanglement wedge $\mathcal{W}_{A}$ and the entanglement wedge cross-section (EWCS)\footnote{Note that the cross-section of the entanglement wedge may have multiple saddle points. In this case the EWCS represents the saddle with the minimal area.} of $\mathcal{W}_{A}$ should change accordingly. Recently the holographic dual of the EWCS attracts considerable attention. There is a zoo of quantum information quantities that are proposed to be the quantum information dual of the EWCS, including the entanglement of purification \cite{Terhal:2002,Takayanagi:2017knl,Nguyen:2017yqw},
the entanglement negativity \cite{Kudler-Flam:2018qjo,Kusuki:2019zsp,Chaturvedi:2016rcn}, the reflected entropy \cite{Dutta:2019gen}, the odd entropy  \cite{Tamaoka:2018ned}, the differential purification \cite{Espindola:2018ozt}, the entanglement distillation \cite{Agon:2018lwq, Levin:2019krg} and balanced partial entanglement entropy (BPE) \cite{Wen:2021qgx,Wen:2022jxr,Camargo:2022mme,Basu:2022nyl}. The BPE has lots of advantages, for example it is easy to compute, can be defined under different purifications, can be defined in non-holographic systems and reduces to the reflected entropy in canonical purification (see the introduction section of \cite{Wen:2022jxr}). The correspondence between the BPE and the EWCS has also passed highly non-trivial tests in both static or covariant configurations in AdS$_3$/CFT$_2$ \cite{Wen:2021qgx,Wen:2022jxr}, in holographic CFT$_2$ with gravitational anomaly \cite{Wen:2022jxr}, and even in 3d flat holography \cite{Bagchi:2010zz,Jiang:2017ecm,Camargo:2022mme,Basu:2022nyl}.

Quite recently, the PEE and BPE were studied in configurations with entanglement islands \cite{Basu:2023wmv}.
By taking into account the island contributions, the authors gave a generalized formula to construct the PEE and compute the BPE. Here the so-called ownerless island region, which lies inside the island Is$(AB)$ of $AB$ but outside Is$(A)\cup$Is$(B)$, plays a crucial role. Remarkably, we find that under different assignments for the ownerless island, we get different BPEs, which exactly correspond to the different saddles for the cross-section of the entanglement wedge $\mathcal{W}_{AB}$. 
The calculations are performed in several set-ups, including the Weyl transformed CFT \cite{Basu:2022crn} (coupled to gravity) and a generalized AdS/BCFT model with a defected theory localized on the brane \cite{Deng:2020ent}. 

In this paper we will test the correspondence between BPE and EWCS in  a modified version of the AdS/BCFT, where the KR (or EoW) brane in the bulk fluctuates. It is interesting that the effective theory for the scalar freedom that characterizes the first order fluctuation of the brane is exactly the JT gravity \cite{Geng:2022slq,Geng:2022tfc,Deng:2022yll} (see also \cite{Bhattacharya:2023drv,Aguilar-Gutierrez:2023tic} for studies with fluctuating KR  brane). We will simulate this configuration using the Weyl transformed CFT$_2$ and adapt the Weyl transformations to the fluctuations of the brane. An import observation we will make is that, unlike the non-fluctuating configuration, the twist operators are inserted at different places respectively in the Weyl-transformed CFT at the asymptotic boundary and the KR brane in the bulk. And only the inserting position on the Weyl-transformed CFT$_2$ will generate the right BPE that coincide with the EWCS. 
This strongly indicates that the KR brane could be emergent from the Weyl transformed CFT$_2$ at the asymptotic boundary.

The paper is organized as follows. 
In Sec.\ \ref{sec:intro-pee}, we start with a brief introduction to the concept of PEE and BPE, and their construction in island phases.
In Sec.\ \ref{sec:revist-jt}, we will first give a quick review on the JT gravity from fluctuating Karch-Randall brane in the AdS/BCFT set-up.
Then the main task of this section is devoted to understanding why the island boundary point no longer corresponds to the intersection point between the RT-like surface and the brane in light of Weyl transformed CFT.
Sec.\ \ref{sec:ew-jt} and Sec.\ \ref{sec:bpe-jt} are devoted to study the correspondence between BPE and EWCS in the AdS/BCFT set-up or Weyl transformed CFT set-up with a fluctuating brane.
We will start with the derivation of the area of EWCS in Sec.\ref{sec:ew-jt}.
We conclude our results in Sec.\ref{sec:dis}.

\section{Backgrounds on PEE and BPE in island phases, and holographic Weyl transformed CFT$_2$}\label{sec:intro-pee}
In this section, we give a brief introduction to PEE and BPE, and their construction in island phases \cite{Basu:2023wmv}.
For the union of sets, we will suppress "$\cup$" symbol for simplicity, i.e. we define $AB\equiv A\cup B $.

\subsection{Partial entanglement entropy}
The partial entanglement entropy (PEE) \cite{Chen:2014,Wen:2018whg,Kudler-Flam:2019oru,Han:2019scu,Wen:2019iyq,Wen:2020ech,Han:2021ycp}\footnote{See \cite{Ageev:2021ipd,Rolph:2021nan,Lin:2022aqf,Lin:2023orb,Liu:2023djf,Lin:2023rbd} for other recent progress on PEE.} is a two-body measure of entanglement featured by the property of additivity. 
It is usually written as $\mathcal{I}(A,B)$ for two non-overlapping spacelike separated subsystems $A$ and $B$. This is the \textit{two-body correlation representation} for the PEE. Note that, we should not mix between the PEE $\mathcal{I}(A,B)$ and the mutual information $I(A,B)$. Although the explicit definition for the PEE based on reduced density matrix is still not clear, in many configurations the PEE can be determined by its physical requirements, which include all the requirements satisfied by the mutual information $I(A,B)$ and the additional requirement of additivity \cite{Chen:2014,Wen:2019iyq}, i.e.
\begin{enumerate}
\item
\textit{Additivity:} $\mathcal{I}(A,BC)=\mathcal{I}(A,B)+\mathcal{I}(A,C)$;

\item
\textit{Permutation symmetry:} $\mathcal{I}(A,B)=\mathcal{I}(B,A)$;

\item
\textit{Normalization:} $\mathcal{I}(A,\bar{A})=S_{A}$;
\item
\textit{Positivity:} $\mathcal{I}(A,B)>0$;
\item
\textit{Upper bounded:} $\mathcal{I}(A,B)\leq \text{min}\{S_{A},S_{B}\}$;
\item
\textit{$\mathcal{I}(A,B)$ should be Invariant under local unitary transformations inside $A$ or $B$};
\item
\textit{Symmetry:} For any symmetry transformation $\mathcal T$ under which $\mathcal T A = A'$ and $\mathcal T B = B'$, we have $\mathcal{I}(A,B) = \mathcal{I}(A',B')$.
\end{enumerate}
It has been demonstrated that \cite{Casini:2008wt,Wen:2019iyq}, the above requirements have a unique solution for Poincar\'e invariant theories, and furthermore, for conformal field theory the formula of the solution can be determined by the requirements. There are also other prescriptions to construct the PEE satisfying the above requirements, like the Gaussian formula applies for Gaussian states \cite{Chen:2014}, geometric construction applied to holographic configurations with a local modular Hamiltonian \cite{Wen:2018whg,Wen:2020ech}, and the so-called additive linear combination (ALC) proposal \cite{Wen:2018whg,Wen:2020ech} applied to generic systems settled on a chain (or line) with a definite order. The ALC proposal will be the workhorse in this paper, which is given in the following.
\begin{itemize}
\item \textit{The ALC proposal \cite{Wen:2018whg,Wen:2020ech,Wen:2019iyq}}: 

Consider a boundary region $A$ and partition it into three non-overlapping subregions $A=\alpha_L\cup\alpha\cup\alpha_R$, where $\alpha$ is some subregion inside $A$ and $\alpha_{L}$ ($\alpha_{R}$) denotes the regions left (right) to it. On this configuration, the claim of the \textit{ALC proposal} is the following:
\begin{align}\label{alc}
s_{A}(\alpha)=\mathcal{I}(\alpha,\bar{A})=\frac{1}{2}\left(S_{ \alpha_L\cup\alpha}+S_{\alpha\cup \alpha_R}-S_{ \alpha_L}-S_{\alpha_R}\right)\,. 
\end{align}
\end{itemize}
By additivity and permutation symmetry, the PEE $\mathcal{I}(A,B)$ can be decomposed into two-point PEEs $\mathcal{I}(\textbf{x},\textbf{y})$ , i.e.
\begin{equation}
\mathcal{I}\left(A, B\right)=\int_{A} \D \sigma_{\textbf{x}} \int_{B} \D \sigma_{\textbf{y}}~ \mathcal{I}(\textbf{x}, \textbf{y})\,,\quad \textbf{x}\in A\,,\quad \textbf{y}\in B\,,
\end{equation}
which means all the PEEs can be generated from the two point PEEs. In a discrete chain, the number of the two-point PEEs equals to the number of all connected intervals, which means that the information contained in all the two-point PEEs is indeed a reformulation of the information of all the entanglement entropy for connected intervals. 

The concept of PEE originates from the study of the entanglement contour $s_{A}(\textbf{x})$ \cite{Chen:2014}, which is assumed to be a spatial density function of entanglement entropy that captures the contribution from each local degree of freedom to $S_A$. The quantity $s_{A}(A_i)$ is then defined as the collections of the entanglement contributions from a subregion $A_i$ of $A$ to the entanglement entropy $S_{A}$, i.e. 
\begin{equation}
    s_{A}(A_i)=\int_{A_i}s_A(\vec x)\D\sigma_{\vec x},
\end{equation}
where $\sigma_{\vec x}$ is an infinitesimal subset. This is just the \textit{contribution representation} of the PEE,
\begin{equation}\label{2-c1}
    \mathcal{I}(A_i,\bar A)\equiv s_{A}(A_i),
\end{equation}
where $\bar A$ denotes the complement of $A$ that purifies $A$.

\subsection{Balanced partial entanglement}
The balanced partial entanglement entropy (BPE) is a PEE that satisfies the balanced requirements. For a system $A\cup B$ in a mixed state $\rho_{AB}$, one can introduce an auxiliary system $A'B'$ to purify $AB$ such that the whole system $ABA'B'$ is in a pure state $|\varphi\rangle$ and,
\begin{align}
\text{Tr}_{A'B'}\ket{\varphi}\bra{\varphi}=\rho_{AB}.    
    \end{align}
The pure state on $ABA'B'$ is called a purification of $\rho_{AB}$, which could be highly non-unique. Let us consider the simple examples described in Fig.\ref{fig:bpe}, for each purification the balanced requirements are summarized in the following \cite{Wen:2021qgx}:
\begin{figure}
\centering
\includegraphics[width=0.8\textwidth]{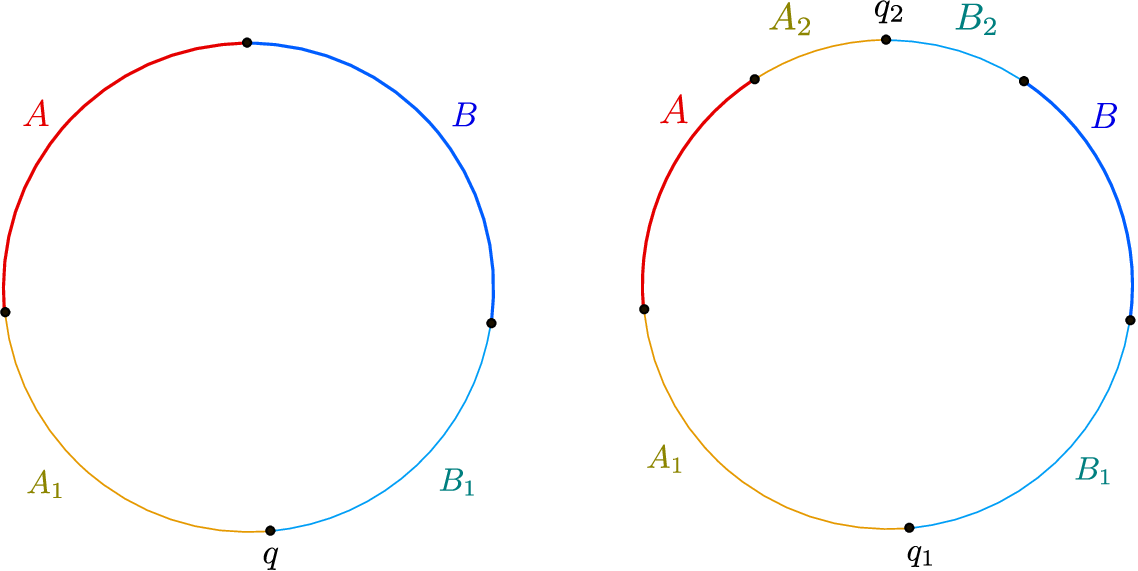}
\caption{
BPE for adjacent and disjoint cases are shown on the left and right respectively. 
$q,q_1,q_2$ are balanced points.
This figure is extracted from \cite{Basu:2023wmv}.}
\label{fig:bpe}
\end{figure}
\subsubsection*{Balance requirements:}
\begin{itemize}
    \item 
When $A$ and $B$ are adjacent, the contribution from $A$ to $S_{AA_1}$ should be equal to the contribution from $B$ to $S_{BB_1}$, that is 
\begin{equation}\label{bc1}
    s_{AA_1}(A)=s_{BB_1}(B),\ \text{or}\quad 
\mathcal{I}(A,BB_1)=\mathcal{I}(B,AA_1).
\end{equation}
This balance requirement is sufficient to determine the partition point in the purifying system $A_1B_1$, which we call the balance point. 

\item When $A$ and $B$ are non-adjacent, the balance requirements become 
\begin{equation}\label{bc2}
    s_{AA_1A_2}(A_1)=s_{BB_1B_2}(B_1),\ s_{AA_1A_2}(A)=s_{BB_1B_2}(B),
\end{equation}
or 
\begin{equation}  
    \mathcal{I}(A,B_1B_2)=\mathcal{I}(B,A_1A_2),\ \mathcal{I}(A_1,BB_2)=\mathcal I(B_1,AA_2),
\end{equation}
which are sufficient to determine the two partition points of the purifying system $A_1B_1A_2B_2$.  Since $S_{AA_1A_2}=S_{BB_1B_2}$, $s_{AA_1A_2}(A_2)=s_{BB_1B_2}(B_2)$ is automatically satisfied provided the satisfaction of the above requirements.  
\item
Note that, it is possible that the solution to the balance requirements is non-unique. In such cases, we should choose the solution that minimizes the BPE, which will be defined soon.
\end{itemize}

{ Provided the balanced requirements are fulfilled, the BPE is then defined as 
\begin{align}\label{BPEd}
	\begin{cases}
		\textit{adjacent cases}:\, &\text{BPE}(A,B)=s_{AA_1}(A)|_{\rm balanced}=s_{BB_1}(B)|_{\rm balanced}\\ \\
	\textit{non-adjacent cases}:\, &\text{BPE}(A,B)=s_{AA_1A_2}(A)|_{\rm balanced}=s_{BB_1B_2}(B)|_{\rm balanced}\,,
	\end{cases} 
\end{align}
Since we have $\mathcal{I}(A,B')=\mathcal{I}(A',B)$ at the balance point\footnote{Here $A'=A_1$ ($B'=B_1$) for adjacent configurations, and $A'=A_1A_2$ ($B'=B_1B_2$) for non-adjacent configurations.}, the BPE can also be expressed as
\begin{equation}
    \text{BPE}(A,B)
    =\mathcal{I}(A,B)+\frac{(\mathcal{I}(A,B')+\mathcal{I}(A',B))|_{\rm balanced}}{2}.
\end{equation}}
It is important to note that \cite{Camargo:2022mme}, the summation $\mathcal{I}(A,B')+\mathcal{I}(A',B)$ which we call the crossing PEE is minimized at the balanced point. This indicates that the BPE can be defined via an optimization problem. It is also interesting to note that, when $A$ and $B$ are adjacent, this minimal crossing PEE gives half of the lower bound of Markov gap \cite{Zou:2020bly,Wen:2021qgx,Hayden:2021gno}, which is a universal constant $(c/6)\log 2$.

\subsection{Ownerless island and generalized ALC formula}\label{sec:intro-pee-is}
Now we generalize our construction of the PEE and BPE to configurations with entanglement islands \cite{Basu:2023wmv}. Here we just list the basic elements we need to carry out the computations. One should consult \cite{Basu:2023wmv} for more details about the physical reasoning behind our generalization. Let us start with the region $AB$ and the island regions Is$(AB)$, Is$(A)$ and Is$(B)$. According to \cite{Basu:2023wmv}, the degrees of freedom in $\is (AB)$ is not independent, and contributes to the entanglement entropy $S_{AB}$ calculated by the island formula. The PEE, for example $s_{AB}(A)$, should contain the contribution from the island region, and it is clear that we should assign the contribution from $\is (A)$ to  $s_{AB}(A)$ and similarly assign $\is (B)$ to  $s_{AB}(B)$. Nevertheless, there are scenarios with regions included in $\is (AB)$ but outside $\is(A)\cup\is(B)$, which should also be assigned to $s_{AB}(A)$ or $s_{AB}(B)$. In other words, when $\is (AB)\neq \emptyset$ and $\is (AB)\supset \is (A)\cup \is (B)$, we define the ownerless island region to be
     \begin{align}
    \text{Io}(AB) =\is(AB)/(\is (A)\cup \is (B))\,.
     \end{align}
The ownerless island should be further divided into two parts
\begin{align}
  \text{Io}(AB)=\text{Io}(A)\cup \text{Io}(B)\,,
  \end{align}
which are assigned to $s_{AB}(A)$ and $s_{AB}(B)$ respectively. We will specify the division of Io$(AB)$ later. All in all, we define the  \emph{generalized islands}
\begin{align}\label{genisland}
 \mathrm{Ir}(A)=\is(A)\cup \text{Io}(A),\qquad \mathrm{Ir}(B)=\is(B)\cup \text{Io}(B)\,,
 \end{align}
 and assign them to $A$ and $B$ respectively when calculating the PEEs. We can classify the assignment into three classes.
\begin{itemize}
    \item 
    \textbf{Class 1:}
    When Is$(AB)$=Is$(A)$=Is$(B)$=$\emptyset$, we have
    \begin{align}
     \mathrm{Ir}(A)=\mathrm{Ir}(B)=\emptyset\,.
     \end{align}  
    \item
    \textbf{Class 2:}
     When Is$(AB)$=$\is(A)\cup \is (B)\neq\emptyset$, we have 
     \begin{align}
     \text{Io}(A)=\text{Io}(B)=\text{Io}(AB)=\emptyset, \qquad \mathrm{Ir}(A)=\is(A),\qquad \mathrm{Ir}(B)=\is(B)\,.
     \end{align}
          \item 
     \textbf{Class 3:}
When we have non-trivial ownerless island region, then we have \eqref{genisland}.
\end{itemize}

Now we turn to the computation of the PEEs and denote $C \equiv\overline{AB\cup\text{Is}(AB)}$ for convenience. After taking into account the contributions from the islands, we should have \cite{Basu:2023wmv},
\begin{equation} \label{pee0}
    s_{AB}(A)=\mathcal{I}(A\cup\text{Ir}(A),C),\quad 
    s_{AB}(B)=\mathcal{I}(B\cup\text{Ir}(B),C).
\end{equation}
A key step to compute the PEEs \eqref{pee0} is to write the right hand side of \eqref{pee0} as a linear combination of the PEEs that can be written in this form $\mathcal{I}(\gamma,\bar\gamma)$, i.e. a PEE between a region $\gamma$ and its complement. For example, using the property of additivity, the PEE \eqref{pee0} is just given by
\begin{equation}\label{galc1}
    \begin{split}
      s_{AB}(A)
      =& \mathcal{I}(A\cup\text{Ir}(A),C)\\
      =& \frac{1}{2}\left[
\mathcal I(A\text{Ir}(A) B \text{Ir}(B), C) + \mathcal I(A\text{Ir}(A), B \text{Ir}(B) C) - \mathcal I(B \text{Ir}(B),A\text{Ir}(A) C)\right]\\
=& \frac{1}{2}\left[ \tilde S_{A\text{Ir}(A)B\text{Ir}(B)}+\tilde S_{A\text{Ir}(A)}- \tilde S_{B\text{Ir}(B)}\right],
    \end{split}
\end{equation}
where we have used $\text{Ir}(A)\text{Ir}(B)=\text{Is}(AB)$, and the notation $\tilde{S}_{\gamma}\equiv\mathcal{I}(\gamma,\bar{\gamma})$ which will be soon explained later in this subsection.

For the adjacent case shown in Fig.\ref{fig:bpe}, we should calculate $s_{AA_1}(A)$ following \eqref{galc1}. For the non-adjacent case in Fig.\ref{fig:bpe} where $A$ is sandwiched by two regions $A_1$ and $A_2$, we have
\begin{equation}\label{galc2}
	s_{A_1AA_2}(A)=\frac{1}{2}\left[ \tilde S_{A_1\text{Ir}(A_1)A\text{Ir}(A)}-\tilde S_{A_1\text{Ir}(A_1)}+\tilde S_{A\text{Ir}(A)A_2\text{Ir}(A_2)}- \tilde S_{A_2\text{Ir}(A_2)}\right].
\end{equation}
This is our \emph{generalized ALC formula} \cite{Basu:2023wmv} in island phases, which is just the ALC formula \eqref{alc} with the replacement $S_{\gamma}\Rightarrow \tilde S_{\gamma\text{Ir}(\gamma)}$ applied to each term. Accordingly, the balance requirements should also be modified to generalized versions, which are given by
\begin{align}\label{gbc1}
		\textit{adjacent cases}:
		\quad \mathcal{I}(A\,\text{Ir}(A),B_1\,\text{Ir}(B_1))=\mathcal{I}(A_1\,\text{Ir}(A_1),B\,\text{Ir}(B)).		
\end{align}
and
\begin{align}\label{gbc2}
	\textit{non-adjacent cases}:
\begin{cases}
		\mathcal{I}(A_1\text{Ir}(A_1),B\,\text{Ir}(B)\,B_2\text{Ir}(B_2))=\mathcal{I}(B_1\text{Ir}(B_1),A\,\text{Ir}(A)\,A_2\text{Ir}(A_2)), \\ \\
		\mathcal{I}(A\,\text{Ir}(A),B_1\text{Ir}(B_1)\,B_2\text{Ir}(B_2))=\mathcal{I}(B\,\text{Ir}(B),A_1\text{Ir}(A_1)\,A_2\text{Ir}(A_2)).
\end{cases} 
\end{align} 
In summary, in island phases the BPE are still defined by \eqref{BPEd}, but we need to use the generalized versions of the ALC formula and balanced requirements.

Before we compute the BPE, we need to clarify how to compute $\mathcal{I}(\gamma,\bar{\gamma})$. In this paper we need to deal with two types of  $\gamma$, 
\begin{itemize}
	\item $\gamma=[-a,b]$ a connected interval,
	\item $\gamma=A\cup \text{Ir}(A)=[-d,-c]\cup[a,b]$ is consists of two disconnected interval,	
\end{itemize}
where $a,b,c,d>0$. For the above two cases, there are two corresponding proposals \cite{Basu:2023wmv} to compute the PEE respectively,
\begin{itemize}
	\item \textit{Basic proposal 1}: $\mathcal{I}(\gamma,\bar\gamma
	)=\tilde{S}_{\gamma}=\tilde S_{[-a,b]}$,
	\item \textit{Basic proposal 2}: $\mathcal{I}(\gamma,\bar\gamma)=\tilde{S}_{[-d,-c]\cup[a,b]}=\tilde{S}_{[-c,a]}+\tilde{S}_{[-d,b]}$,
\end{itemize}
where, for example, $\tilde S_{[-a,b]}$ is the two-point function of twist operators inserted at $x=-a$ and $x=b$. Such two-point functions are very subtle in the effective theory of AdS/BCFT correspondence or other doubly holography configurations, since the region $[-a,0]$ may not be the island region of $[0,b]$. In other words, the island formula \eqref{island_EE} only involves $\tilde S_{\gamma}$ where $\gamma=\mathcal{R}\cup \text{Is}(\mathcal{R})$ is the union of a region and its island, and in this case $\tilde S_{\gamma}=S_{\mathcal{R}}$. While the $\gamma$ we deal with usually goes beyond this type, and $\tilde S_{\gamma}$ should not be understood as the entanglement entropy of $\gamma$ or any other region (see \cite{Basu:2023wmv} for more discussions). Nevertheless, these two-point functions are well defined in the Weyl transformed CFT, and are computed by applying the Weyl transformation to the two-point functions in normal CFT$_2$ (see \eqref{twopointf})). 

Though the above two basic proposals have not been proved yet, they have produced highly consistent results between the BPE and the EWCS in various configurations with entanglement islands \cite{Basu:2023wmv}. The \textit{basic proposal 1} looks like a generalization of the $RT$ formula in AdS$_3$/CFT$_2$ for single interval. While the \textit{basic proposal 2} is a generalization of the $RT$ formula for two-intervals with connected entanglement wedge. It looks reasonable as this proposal applies when the region $[a,b]$ admits an island, hence the entanglement wedge looks more like the connected phase of a two-interval. Note that the \textit{basic proposal 2} is not consistent with the normalization and additive property of the PEE, which may be explained by a similar phase transition of the PEE flux in AdS$_3$/CFT$_2$\footnote{This phase transition should originate from the large $c$ limit of the holographic CFT$_2$ \cite{Hartman:2013mia}.} \cite{Lin:2023rbd,Lin:2024dho}. We leave this for future investigation \cite{WenXuZhong}. In summary, our calculations in this paper will only involve linear combinations of Weyl transformed two-point functions of the twist operators.

The remaining problem is the division of the ownerless island region $\text{Io}(AB)=\text{Io}(A)\cup\text{Io}(B)$. In \cite{Basu:2023wmv}, the authors considered the AdS/BCFT set-up and its simulation via a holographic Weyl transformed CFT$_2$. They found that different assignments for the ownerless island region lead to different BPEs, which exactly correspond to different saddles of the EWCS. Then according to the minimum requirement, we should choose the assignment that gives us the minimal BPE. It seems that, the division of the ownerless island depends on the phase structure of the EWCS.

\subsection{Simulation of AdS/BCFT correspondence via holographic Weyl transformed CFT$_2$}\label{subsec-simulation}
\begin{figure}
	\centering
	\includegraphics[width=0.9\textwidth]{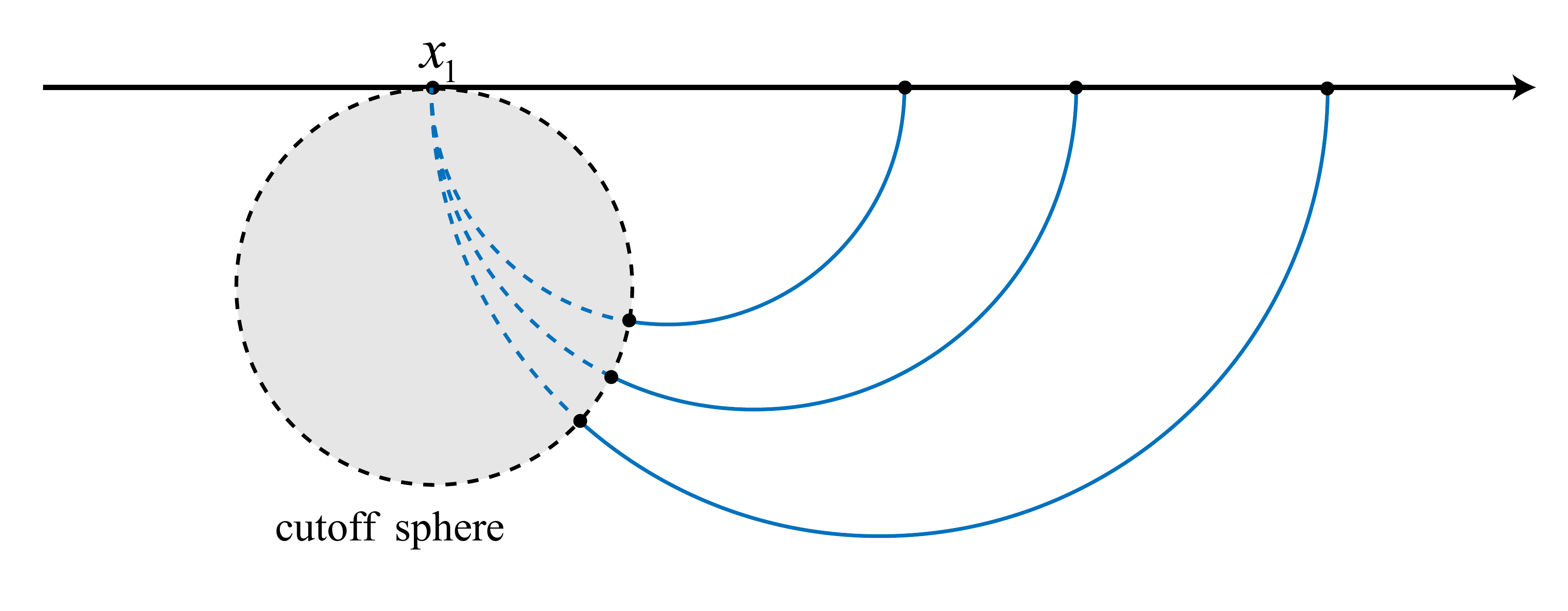}
	\caption{Cutoff sphere for Weyl transformed CFT. Note that the RT curves are always normal to the cutoff sphere.
	}
	\label{fig:cut-sphe}
\end{figure}

The Weyl transformed CFT$_2$ simulation for AdS$_3$/BCFT$_2$ correspondence was first introduced in \cite{Basu:2022crn,Suzuki:2022xwv}. 
Such Weyl transformations depend on the UV cutoff and introduces finite cutoff in certain region. { The Weyl transformation is assumed to be the operation that vastly reduces the Hilbert space of the theory and hence induces the self-coding property and the island formula for entanglement entropy \cite{Basu:2022crn}. Since the scalar field determines the metric in the $x<0$ region, its dynamics should be a gravitational theory. Also, it was assumed that the region with non-trivial Weyl transformation is coupled to an induced gravity, which describes the dynamics of the scalar field which characterizes the Weyl transformation \cite{Basu:2022crn}. Hence the island formula $I$ \eqref{island_EE} applies. } 

Let us start with a holographic CFT$_2$ with UV cutoff $\epsilon$.
According to the AdS/CFT, the metric background of the theory should be $\D s^2=(-\D t^2+\D x^2)/\epsilon^2$. Then we perform a Weyl transformation on the theory, which changes the metric in the following way
\begin{equation}
	\D s^2= \frac{-\D t^2+\D x^2}{\epsilon^2},\quad  
	\Rightarrow\quad 
	\D s^2=e^{2 \varphi(x)}\left(\frac{-\D t^2+\D x^2}{\epsilon^2}\right).
\end{equation}
This effectively changes the cutoff in a position-dependent way,
\begin{equation}
	\epsilon\ \to\ e^{-\varphi(x)}\epsilon.
\end{equation}
Accordingly, the two-point function fo the twist operators settled at the endpoints of an interval, for example $A=[a,b]$, changes in a simple way\footnote{Actually the two-point function of the twist operators may have a contribution from gravitational fluctuations, which is a area term according to \cite{Lewkowycz:2013nqa}. Nevertheless we will ignore this term in the cases we study. } \cite{Caputa:2017urj,Caputa:2018xuf,Camargo:2022mme}
\begin{align}\label{twopointf}
	\tilde{S}_{[a,b]}=\frac{c}{3}\log \frac{(b-a)}{\epsilon}+\frac{c}{6}\varphi(a)+\frac{c}{6}\varphi(b)\,,
\end{align}
which is just the original two-point function in CFT$_2$ adding the value of $\varphi(x)$ at the endpoints. The above formula is the building block of all our calculations in this paper. Holographically, as shown in Fig.\ \ref{fig:cut-sphe}, the constant subjection of the entanglement entropy was understood as inserting cutoff spheres with radius $|\varphi(x)|$ centered at the endpoints \cite{Basu:2022crn}. 
When computing the holographic entanglement entropy, we should exclude the portion of the RT surface inside the cutoff spheres.

The particular Weyl transformation that captures the main features of the AdS/BCFT correspondence (with a non-fluctuating KR brane) was given in\footnote{The scalar field looks not smooth or continues at $x=0$. If one is uncomfortable about this, one can smooth the scalar field at $x=0$, and the calculations in this paper will not be affected by the smoothing operation.} \cite{Basu:2022crn},
\begin{equation}\label{varphi1}
	\varphi(x)= \begin{cases}0, & \text { if } \quad x>0 \\ -\log \left(\frac{2|x|}{\epsilon}\right)+\kappa_1,  & \ \text {if } \quad x<0\end{cases}.
\end{equation}
In this case, the common tangent line (or the envelope surface) of all the cutoff spheres is exactly a straight line locates at
\begin{align}
	\rho=\kappa_1\,.
\end{align}

\begin{figure}
	\centering
	\includegraphics[width=0.7\textwidth]{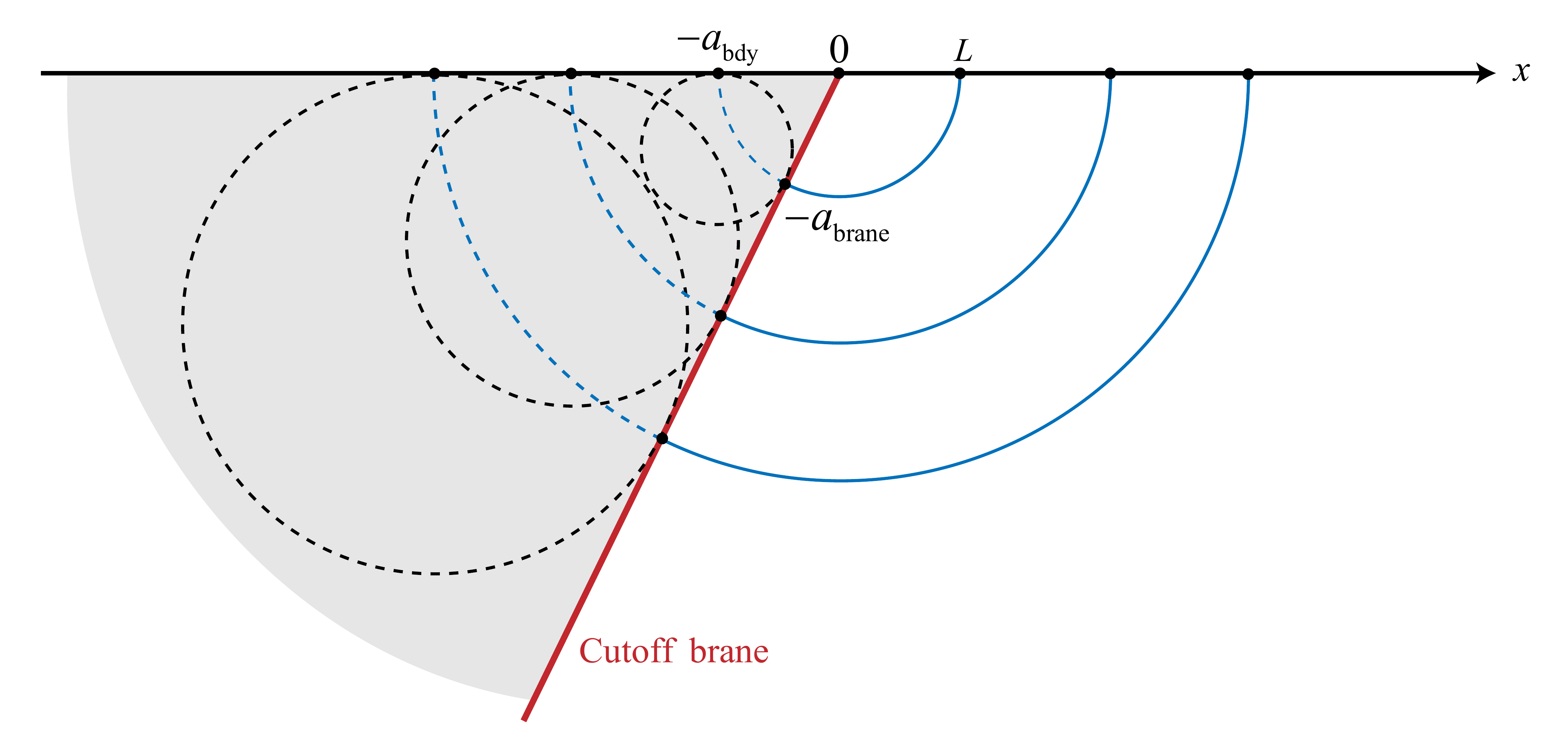}
	\caption{The cutoff spheres and their common tangent line, the cutoff brane, in Weyl transformed CFT$_2$. The cut off brane is captured by the function $\rho=\kappa_1$, which coincide with the KR brane at $\rho=\rho_0$ in AdS/BCFT if we set $\kappa_1=\rho_0$. 
	}
	\label{fig:WCFT1}
\end{figure}

{ In a following-up paper \cite{Chandra:2024bkn}, it was shown that the above scalar field is the one that optimizes the path-integral computation of the reduced density matrix for the $x>0$ region. It is also the saddle point of the Liouville action, which could be the suspected gravitational theory coupled to the $x<0$ region.} Since the $x<0$ region is coupled to the gravity describing the dynamics of the scalar field, the entanglement entropy for regions on the right hand side $x>0$ could enter island phase according to the island formula $I$. For regions in island phase, the RT surfaces are cut off exactly at the cutoff brane, see Fig.\ \ref{fig:WCFT1}. In other words the RT surface is allowed to anchor on the cutoff brane, which plays the same role as the KR brane in the AdS/BCFT correspondence. Since $\Omega(x)^{-2}=e^{2\varphi(x)}$, this explains the choice of the Weyl factor 2 \eqref{Omega2} in the AdS/BCFT set-up from the cutoff point of view.

On the field theory side, let us apply the island formula $I$ to calculate $S_A$ for the interval $A=[0,L]$ in this Weyl transformed CFT and assuming the island region $\text{Is}(A)=[-a,0)$, then we have
\begin{align}
	S_{A}=\text{min~ext}_a~[\tilde{S}_{[-a,L]}]=\text{min~ext}_a \left[\frac{c}{3}\log \frac{(L+a)}{\epsilon}+\frac{c}{6}\varphi(-a)+\frac{c}{6}\varphi(L)\right]\,.
\end{align}
Plugging into \eqref{varphi1}, the above formula is minimized at
\begin{align}
	a_{\rm bdy}=L,
\end{align}
hence
\begin{align}\label{ee-wcft0}
	S_{A}=\frac{c}{6}\log \frac{2L}{\epsilon}+\frac{c}{6}\kappa_1\,.
\end{align}
This exactly matches the result in AdS/BCFT with $\rho_0=\kappa_1$. 

On the gravity side, the holographic entanglement entropy is given by the area of the minimal extreme surface (the RT surface) which is homologous to $A$ and is allowed to be cut off at any of the cutoff spheres. As expected \cite{Basu:2022crn}, the RT surface is just the circle emanating from the endpoint $x=L$ on the boundary and anchored on the cutoff brane at $y=-a_{\rm brane}=-L$\footnote{Here $y$ is the coordinate parameterizing the brane, see \eqref{ycor}.}, see Fig.\ \ref{fig:WCFT1}. In this case we have $a_{\rm bdy}=a_{\rm brane}$. 

The above configuration exactly matches with the AdS/BCFT correspondence given $\rho_0=\kappa_1$. { The key for this simulation is that, we should adjust $\kappa_1$ such that the cutoff brane overlaps with the KR brane in the AdS/BCFT configuration which we simulate. Also, the coincidence between the two set-ups can be straightforwardly checked for the calculation of the entanglement entropy for any interval $A=[a,b]$ in the $x>0$ region, with phase transitions for the RT surfaces. In the next section, we will consider the simulation for a more generic AdS/BCFT set-up, where the KR brane is allowed to fluctuate in the bulk. Similarly the simulation can be achieved by
adjusting the scalar field that characterizes the Weyl transformation, such that the cutoff brane also fluctuates and overlaps with the fluctuating KR brane.}

\section{AdS/BCFT with fluctuating KR brane and its simulation as a Weyl transformed CFT}\label{sec:revist-jt}

\subsection{JT gravity from the fluctuation of the KR brane}
Let us start with a quick look at the idea of JT gravity from dimensional reduction \cite{Geng:2022slq,Geng:2022tfc,Deng:2022yll}. Consider the AdS/BCFT correspondence with the bulk action,
\begin{equation}
\begin{aligned}
I_{\text{AdS}} & =\frac{1}{16 \pi G_N} \int_N \sqrt{-g}(R-2 \Lambda)  +\frac{1}{8 \pi G_N} \int_{\mathcal Q} \sqrt{-h} (K-T),
\end{aligned}
\end{equation}
where $N$ denotes the bulk AdS spacetime and 
$\mathcal Q$ denotes the Karch-Randall (KR) brane with the Neumann boundary condition imposed and $T$ is the brane tension.
It is convenient to write the  bulk metric in terms of the following coordinates,
\begin{align}
\D s^2=&\frac{\ell^2}{z^2}(-\D t^2+\D x^2+\D z^2)\\\label{ycor}
=&\D \rho^2+\ell^2 \cosh ^2 \frac{\rho}{\ell} \frac{-\D t^2+\D y^2}{y^2}\,,
\end{align}
where $\ell$ is the AdS radius, which we will take to be unit $\ell=1$ in the following, and the two sets of coordinates are related by,
\begin{equation}\label{eq:coor-rela}
    z=y\cosh^{-1}{\rho},\quad x=-y\tanh {\rho}.
\end{equation}
Here the brane is located at $\rho=\rho_0$, where $\rho_0$ is a constant determined by the tension of the brane $\rho_0=\text{arctanh}~ T$.

\begin{figure}
    \centering
    \includegraphics[width=0.7\textwidth]{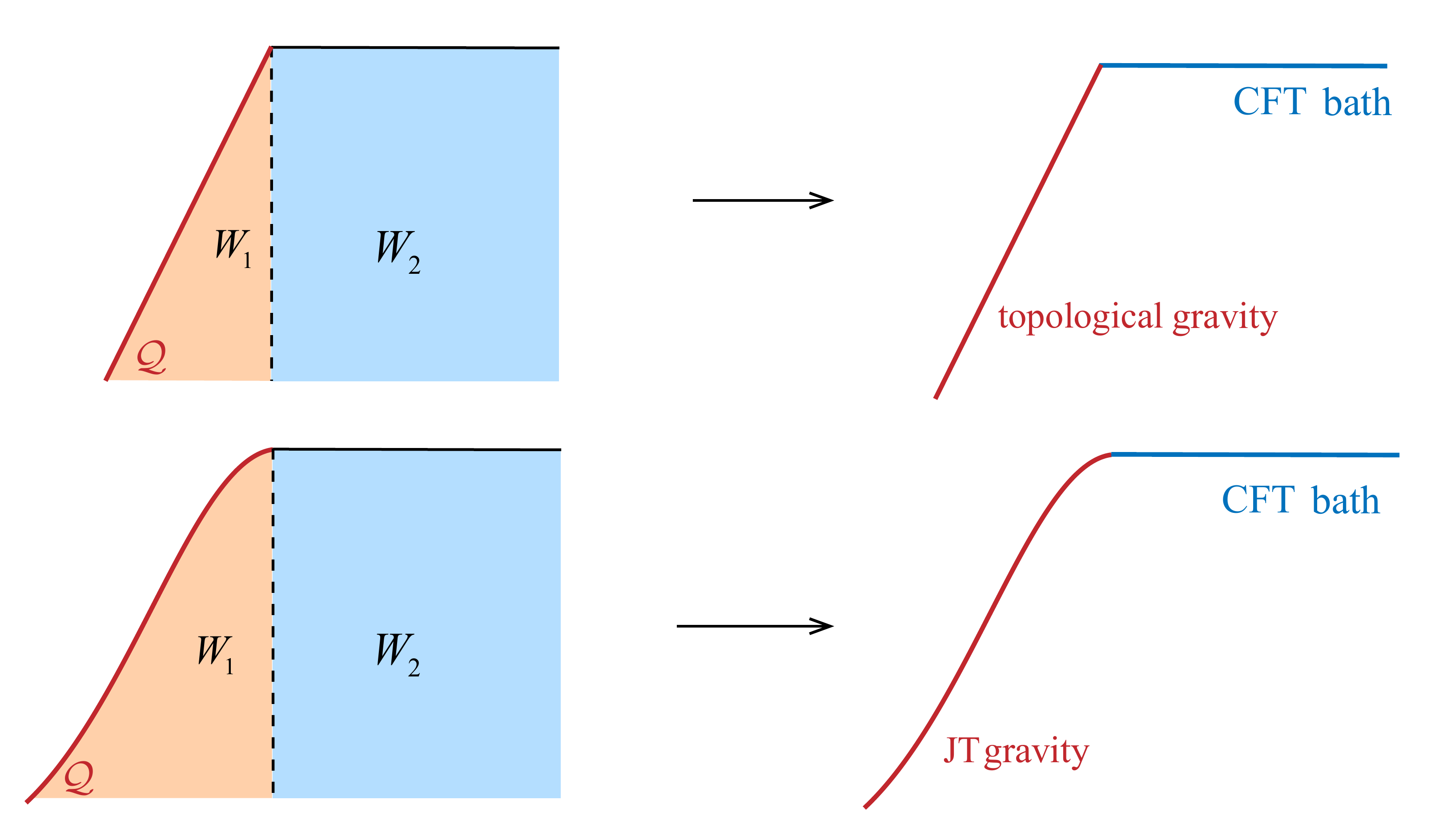}
    \caption{ Gravity on the KR brane $\mathcal Q$ via dimensional reduction on $W_1$.
    The induced gravity on a brane at a fixed $\rho_0$ is topological, while the induced gravity is JT gravity if we allow the brane to fluctuate.
    }
    \label{fig:rd}
\end{figure}

Let us divide the bulk into two wedges $W_1\cup W_2$ along the $\rho=0$  (see Fig.\ \ref{fig:rd}).
By braneworld holography or dimensional reduction \cite{Randall:1999ee,Randall:1999vf,Karch:2000ct} on wedge $W_1$ (see \cite{Deng:2020ent,Suzuki:2022xwv} for example), we will get the induced gravity on the KR brane, which is a topological gravity on the brane (see the upper figure in Fig.\ \ref{fig:rd}). Later we just assume that the gravitational theory on the brane is just the induced gravity. And the region $W_2$ is regarded as the gravity dual of the CFT$_2$ bath. 

Following \cite{Geng:2022slq,Geng:2022tfc,Deng:2022yll}, we let the KR brane  fluctuate around constant $\rho=\rho_0$ slice, and only keep the first order fluctuation. The brane fluctuation $\delta\rho/\rho_0$ can be described by a scalar degree of freedom, and results in a 2$d$ effective action (to the order $\mathcal{O}(\delta\rho/\rho_0)$) on the brane. This theory is exactly the JT gravity coupled to a dilaton characterized by the scalar degree of freedom,
\begin{equation}
\begin{aligned}
I_{2d} & =\frac{\rho_0}{16 \pi G_N} \int \sqrt{-g^{(2)}} \frac{\delta\rho}{\rho_0}\left(R^{(2)}+\frac{2}{ \cosh ^2 {\rho_0}}\right) 
+\frac{\rho_0}{16 \pi G_N} \int \sqrt{-g^{(2)}} R^{(2)}+\mathcal{O}\left(\frac{\delta\rho^2}{\rho_0^2}\right).
\end{aligned}
\end{equation}
The solution of dilaton on the brane ($y<0$) is given by,
\begin{equation}
    \frac{\delta\rho}{\rho_0}=-\frac{\bar\phi_r}{y}\,,\qquad y<0,
\end{equation}
where $\bar\phi_r$ is the renormalized boundary dilaton of JT gravity that is taken as a constant \cite{Maldacena:2016upp}.
Then the brane locates at (see the lower figure in Fig.\ \ref{fig:rd}),
\begin{equation}\label{b-lo}
     \rho=\rho_0+\delta\rho=\rho_0\left(1-\frac{\bar\phi_r}{y}\right).
\end{equation}
Since the induced gravity on the fluctuating brane is JT gravity, later we will also refer to this brane as the JT brane.

\subsection{Entanglement entropy with fluctuating KR brane}
Now we calculate the entanglement entropy in the $2d$ effective theory for an interval $A=[0,L]$ via the island formula $I$. Assuming the island region to be $\is (A)=[-a,0)$, the island formula gives us
 \begin{equation}
   S_{A}=\text{min~ext}_a~[\tilde{S}_{[-a,L]}]=\text{min~ext}_a \left[\frac{\text{Area}(\partial \is(A))}{4G_2}+S_{\text{bulk}}([-a,L])\right]
\end{equation} 
In the AdS/BCFT set-up, there is a holographic picture for the $2d$ effective theory, where the bulk entanglement entropy $S_{\text{bulk}}([-a,L])$ can be calculate by the area of the RT surface $\Gamma$ that anchors on the JT brane (see Fig.\ref{fig:des}). Hence the island formula becomes\footnote{Here the $S_{\text{bulk}}([-a,L])$ is calculated in a semi-classical way thus the bulk entanglement entropy in AdS$_3$ is ignored.}
\begin{align}
S_{A}= \text{min~ext}\left[\frac{\text{Area}(\partial \is(A))}{4G_2}+\frac{\operatorname{Area}(\Gamma)}{4 G_3}\right]
\end{align}
{ In this paper, since we take the full JT gravity on the brane to be induced from the KR reduction of the region $W_1$, the area term can be ignored \cite{Suzuki:2022xwv,Geng:2020fxl,Geng:2022tfc}, hence
\begin{align}
	\tilde{S}_{[-a,L]}=S_{\text{bulk}}([-a,L])\,,
\end{align}
which is calculated by the Weyl transformed two-point function of the twist operators \eqref{twopointf}.} On the effective theory side, $S_{\text{bulk}}([-a,L])$ is calculated by a Weyl transformed two-point function of twist operators. So far, there are two choices of the Weyl transformation, which give different $S_{\text{bulk}}([-a,L])$.

\subsubsection*{Weyl factor 1}
The first one is the Weyl transformation that relates the induced metric on the brane with the flat metric.
Up to the order $\mathcal{O}(\delta\rho)$, the induced metric of the JT brane is given by,
\begin{equation}
\begin{split}
    \D s^2_{\text{JT-brane}}
    &={\cosh^2\rho}\left(\frac{-\D t^2+\D y^2}{y^2}\right)\\
    & \approx 
    \left({\cosh^2\rho_0{+\delta\rho\sinh (2\rho_0)}}\right)\left(\frac{-\D t^2+\D y^2}{y^2}\right)+\mathcal{O}(\delta\rho^2)\\
    &= \Omega(y)^{-2}\frac{-\D t^2+\D y^2}{\epsilon^2}+\mathcal{O}(\delta\rho^2),
    \end{split}
\end{equation}
which is conformal flat in $(t,y)$ coordinates with the conformal factor { 
\begin{equation}\label{Omega1}
    \Omega^{-2}(y)=\frac{\epsilon^2}{y^2} \left({\cosh^2\rho_0+\delta\rho\sinh (2\rho_0)}\right)=e^{2\varphi(y)}.
\end{equation}}
Before we go ahead, we introduce a parameter
\begin{align}
\mu\equiv \frac{\rho_0\bar\phi_r}{6 L}=-\frac{\delta\rho}{6}\frac{y}{L}.
\end{align}
which represents the first order fluctuation of the brane as long as $(y/L)\sim \mathcal O(1)$. In the rest of this paper we will only keep the first order of the fluctuation, and equations with ``$\approx$'' only hold up to the order $\mathcal{O}(\mu)$. 
Note that, compared with previous literature \cite{Almheiri:2019qdq,Deng:2022yll}, we have introduced the additional parameter $\epsilon$ which represents the UV cutoff of the boundary. 
Then the bulk entropy is given by
  \begin{equation}
 \begin{split}
S_{\text{bulk}}([-a,L])&=\frac{c}{6} \log \frac{(L+a)^2 }{ \epsilon^2}+\frac{c}{6}\varphi(-a)
   \\
   &\approx \frac{c}{6} \log \frac{(L+a)^2 }{2 a \epsilon}+\frac{c}{6}\log \left(2 \cosh\rho_0\right)+\frac{cL\mu}{a}\tanh\rho_0,
    \end{split}
\end{equation}
Which is just a Weyl transformed two-point function\footnote{Note the exchange between the $x$ and $y$ coordinates, as the JT gravity lives on the brane which is parameterized by $y$.}. The same calculation was also carried out in \cite{Almheiri:2019qdq} for a non-fluctuating brane, which can be reproduced by our calculation when setting $\delta \rho=0$.

Note that the subtle cutoff scale $\epsilon_{y}$ on the brane (see \cite{Almheiri:2019qdq,Deng:2020ent}) is terminated by the Weyl factor $\Omega(-a)$ as \eqref{Omega1} depend on $\epsilon$ and diverges as $\epsilon\to 0$. The extremal point is given by
\begin{equation}\label{is-a-1}
   a\approx a_{\rm bdy}= L+12 L\mu\tanh\rho_0,
\end{equation}
hence the entanglement entropy from the island formula $I$ is
\begin{equation}\label{Omega1Sa}
    S_{A}\approx \frac{c}{6}\log\frac{2L}{\epsilon}+\frac{c}{6}\log \left(2 \cosh\rho_0\right)+c\mu\tanh\rho_0.
\end{equation}

\subsubsection*{Weyl factor 2} 
The another choice for the Weyl factor (see \cite{Suzuki:2022xwv,Basu:2022crn,Basu:2023wmv} for examples) is motivated from the role of the KR brane as a cutoff boundary of the AdS bulk. The corresponding Weyl factor is given by (see next subsection)
\begin{align}\label{Omega2}
\Omega^{-2}(x)&=\exp\left[
{-2\left(\log \left(\frac{2|x|}{\epsilon}\right)-\rho_0+\frac{\rho_0 \bar{\phi}_r}{x}\right)}
\right]\\
&=\exp\left[
-2\left(
\log\frac{-2x}{\epsilon}-\rho_0+\frac{6\mu L}{x}
\right)
\right]=e^{2\varphi(y)}
 \quad y<0\,,
\end{align}
which results in
  \begin{equation}
 \begin{split}
S_{\text{bulk}}([-a,L])&=\frac{c}{6} \log \frac{(L+a)^2 }{\epsilon^2}+\frac{c}{6}\varphi(-a)
   \\
   &\approx \frac{c}{6}\log\frac{(a+L)^2}{2 a \epsilon}+\frac{c}{6}\rho_0+\frac{cL\mu}{a}\,.    \end{split}
\end{equation}
In this case, the extremal point is given by
\begin{align}\label{is-a-2}
a\approx a_{\rm bdy}=L+12L \mu\,,
\end{align}
hence the entanglement entropy from the island formula $I$ becomes
\begin{align}\label{Omega2Sa}
S_{A}\approx \frac{c}{6}\log\frac{2L}{\epsilon}+\frac{c}{6} \rho_0+c\mu\,.
\end{align}

\begin{figure}
    \centering
    \includegraphics[width=0.55\textwidth]{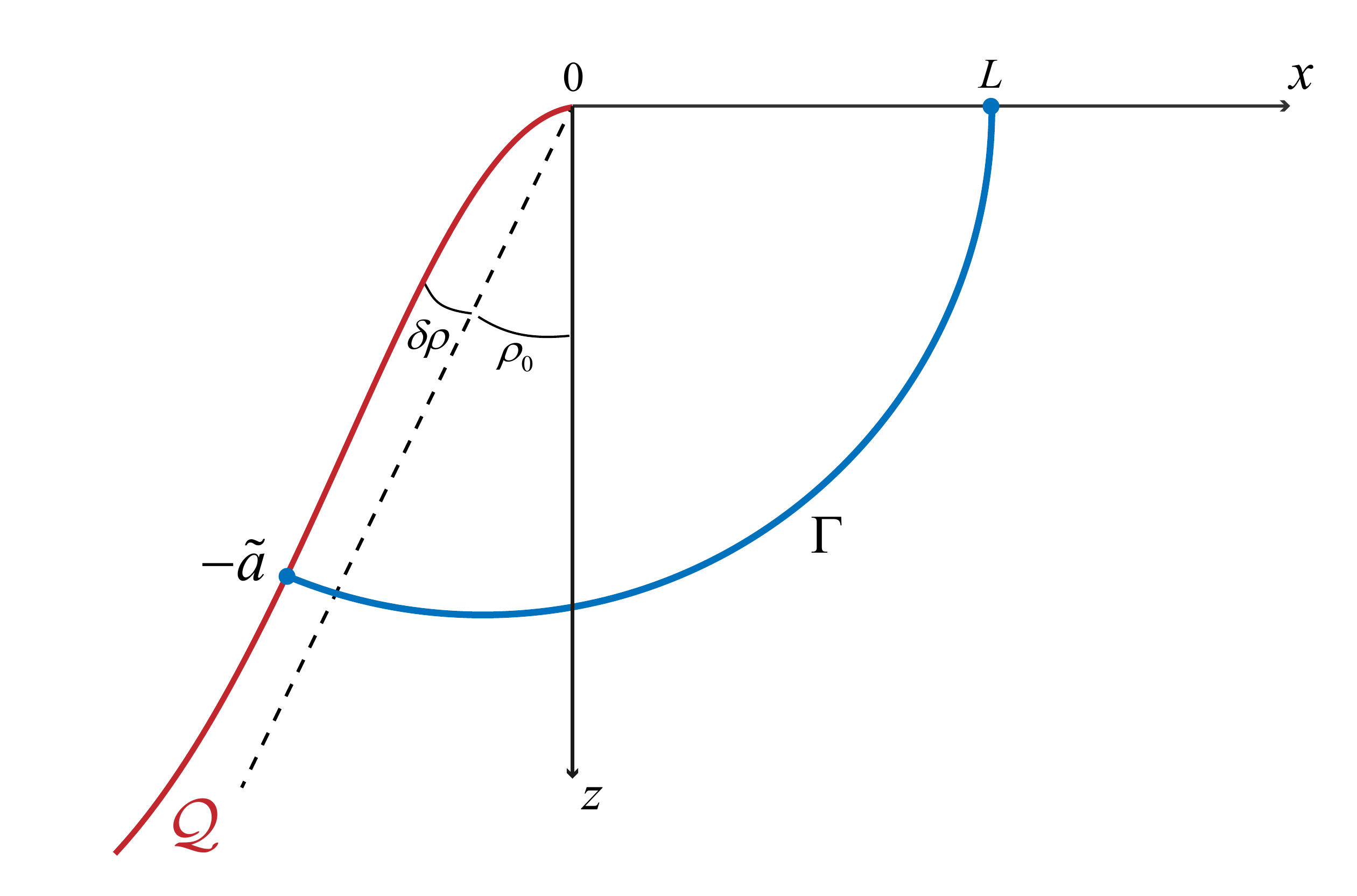}
    \caption{Entanglement entropy for $[0,L]$ from $3d$-bulk RT prescription.
    }
    \label{fig:des}
\end{figure}

\subsubsection*{Doubly holography calculation} 
From the $3d$-bulk viewpoint, the bulk entanglement entropy term in the island formula $I$ is just given by the area of the RT surface $\Gamma$ (see Fig.\ \ref{fig:des}) that anchored on the brane vertically,
\begin{equation}
S_{A}=\frac{\operatorname{Area}(\Gamma)}{4 G_N}.
\end{equation}
The RT surface can be determined by minimizing the length of the geodesic (see the blue curve in Fig.\ \ref{fig:des}) that anchors on the brane at $y=-\tilde{a}$, i.e. \cite{Deng:2022yll}
\begin{equation}\label{RT}
\begin{aligned}
\frac{\operatorname{Area}(\Gamma)}{4 G_N}
&=\text{Min} \left[\frac{c}{6} \log\left(\frac{\left(L+\tilde a \tanh\rho_0\right)^2+\tilde a^2 \cosh ^{-2}\rho_0}{ \tilde a \epsilon\cosh ^{-1}\rho_0 }\right) +\frac{cL \mu}{\tilde a}\right].
\end{aligned}
\end{equation}
The extremal point is given by
\begin{equation}\label{RTsaddle}
 \tilde{a}\approx a_{\rm brane}= L+6 L\left(1+\tanh \rho_0\right)\mu,
\end{equation}
where the subscript ``brane'' means that the extremal point is the point between where the RT-like surface $\Gamma$ anchors on the KR brane.
Finally, the entanglement entropy is
\begin{equation}\label{doublehsa}
    S_{A}\approx\frac{c}{6}\log\frac{2L}{\epsilon}+\frac{c}{6} \rho_0+c\mu.
\end{equation}

\subsubsection*{Two important observations} 
\begin{itemize}
\item Firstly, it is obvious that, the holographic calculation \eqref{doublehsa} only matches entanglement entropy \eqref{Omega2Sa} under the second Weyl transformation \eqref{Omega2}. While the entanglement entropy \eqref{Omega1Sa} under the first Weyl transformation does not match \eqref{doublehsa} even under the non-fluctuating limit $\mu\to 0$. It only match \eqref{doublehsa} in the limit that the KR brane approaches the asymptotic boundary, such that the cutoff $\epsilon_y$ on the brane is satisfy $0<\epsilon\ll\epsilon_y \ll 1$ (see \cite{Almheiri:2019qdq}).

\item Secondly, the minimal point $a_{\rm brane}$ \eqref{RTsaddle} for the RT curve differs from both of the minimal points \eqref{is-a-1} and \eqref{is-a-2} of the island formula, i.e.
\begin{align}
a_{\rm brane}\neq a_{\rm bdy}\,.
\end{align}
This is quite different from the case with non-fluctuating brane, where $a_{\rm brane}=a_{\rm bdy}$ for both of the Weyl factors under the limit $\mu \to 0$.
\end{itemize}
These issues will be well explained based on the simulation for the AdS/BCFT via a Weyl transformed CFT$_2$. { One may think that the $\mathcal{O}(\epsilon^0)$ term is not universal as it depends on the regulator $\epsilon$, hence the difference between our results at order $\mathcal{O}(\epsilon^0)$ is not important. We stress that, firstly in the configurations under consideration the $\mathcal{O}(\epsilon^0)$ order term in the entanglement entropy represents the so-called boundary entropy, which has a physical meaning and does not depend on $\epsilon$.  Secondly, when we compare the results \eqref{Omega1Sa},\eqref{Omega2Sa} and \eqref{doublehsa} we have assumed that they have the same UV cutoff $\epsilon$, hence adjusting the regulator respectively is not allowed. At last, when calculating the PEE or BPE later, we will encounter linear combinations of the entanglement entropies which are independent from the regulator, and the  $\mathcal{O}(\epsilon^0)$ order terms play the essential role.}

\subsection{KR brane vs the cutoff brane from Weyl transformations}\label{sec:wtcft}
\begin{figure}
    \centering
    \includegraphics[width=0.7\textwidth]{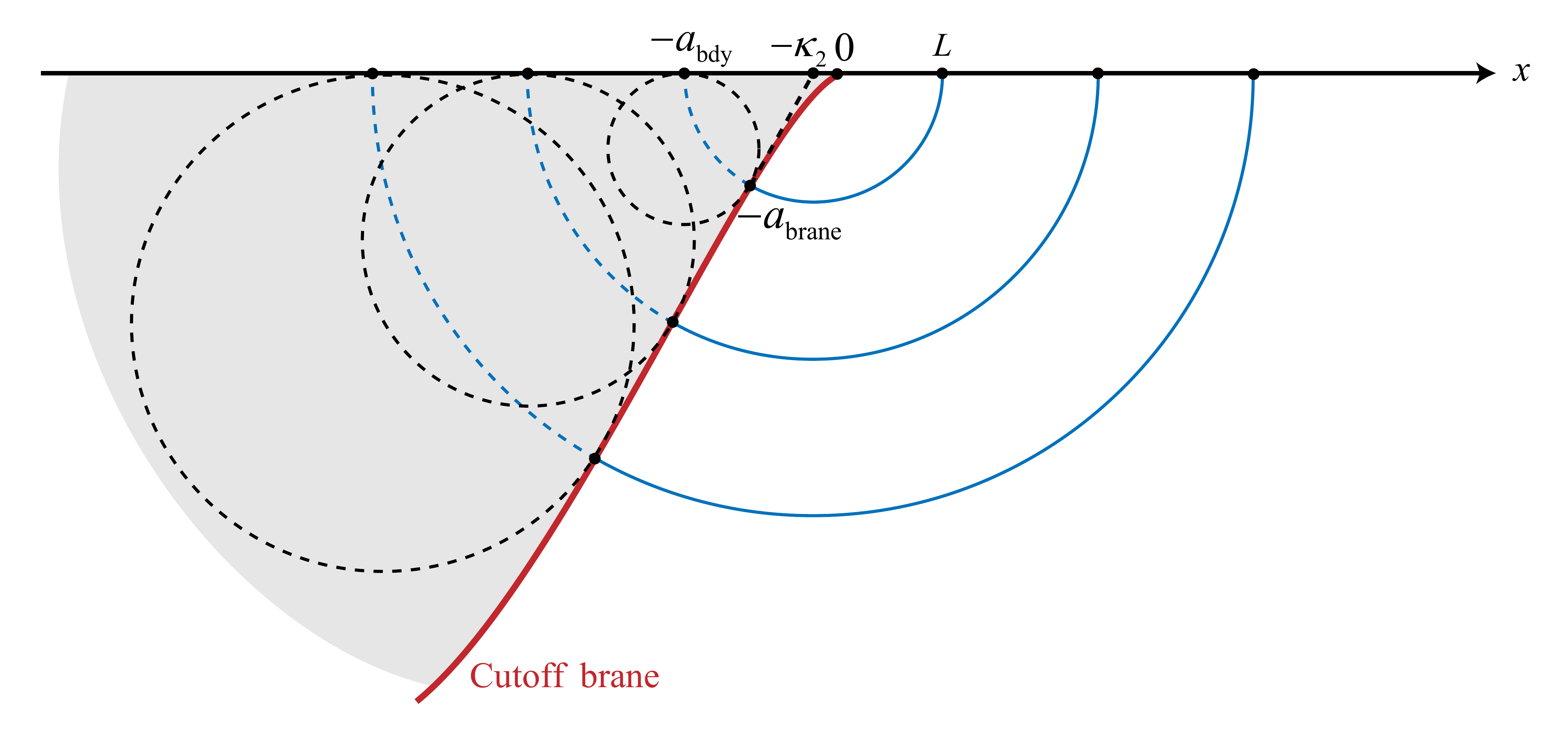}
    \caption{The cutoff brane in Weyl transformed CFT vs the fluctuating KR brane in AdS/BCFT. The dashed part of the blue curves is cut off when computing the area of RT surface.
    }
    \label{fig:WCFT2}
\end{figure}

Now we simulate the AdS/BCFT configuration with fluctuating brane via the Weyl transformed CFT$_2$ with the Weyl transformations modified accordingly. Using \eqref{eq:coor-rela} and \eqref{b-lo}, the location of JT brane (or fluctuating KR brane) is given by the following parametric equation,
\begin{equation}\label{jt-locate-1}
    \begin{split}
        -x_J&=-|y|\tanh\left[ \rho_0\left(1+\frac{\bar\phi_r}{|y|}\right)\right]
        \approx 
        -|y|\tanh  \rho_0-
        {\rho_0\bar\phi_r}\cosh^{-2}\left( \rho_0\right)+\mathcal{O}(\bar\phi_r^2),\\
        z_J&=|y|\cosh^{-1}\left[ \rho_0\left(1+\frac{\bar\phi_r}{|y|}\right)\right]
        \approx
        |y| \cosh^{-1}  \rho_0-
        {\rho_0\bar\phi_r}\cosh^{-1}\left( \rho_0\right)\tanh\left( \rho_0\right)+\mathcal{O}(\bar\phi_r^2).
    \end{split}
\end{equation}

{ In order to simulate the above AdS/BCFT configuration, we should adjust the Weyl transformation such that the corresponding cutoff brane (i.e. the common tangent curve of all the cutoff spheres) overlaps with the JT brane \eqref{jt-locate-1}. Here we first give a guess for the scalar field, then we check that by properly adjusting the parameters in the scalar field, the corresponding cutoff brane can overlap with the JT brane \eqref{jt-locate-1}. It is reasonable to expect that the scalar field is \eqref{varphi1} plus a fluctuation term, which is proportional to the solution of the JT gravity. More explicitly, it can be written as, 
\begin{equation}\label{varphi2}
\varphi(x)= \begin{cases}0, & \text { if } \quad x>0 \\ -\log \left(\frac{2|x|}{\epsilon}\right)+\kappa_1-\frac{\kappa_2}{x},  & \ \text {if } \quad x<0\end{cases}.
\end{equation}
As we have mentioned at the beginning of section \eqref{subsec-simulation}, we assumed the fluctuation of the scalar field is described by a gravitational theory coupled to the Weyl transformed region, and according to \cite{Geng:2022slq,Deng:2022yll} this gravitational theory should be the JT gravity. Then the solution to the JT gravity is a natural guess for the fluctuation of the scalar field.} The corresponding cutoff brane is then parameterized by (see Appendix.\ \ref{app:cut} for details)
\begin{equation}
\begin{split}
    -x_m& \approx -|y|\tanh{\kappa_1}-\kappa_2\cosh^{-2}\kappa_1+\mathcal{O}(\kappa_2^2),\\
    z_m& \approx
    |y|\cosh^{-1}{\kappa_1}-\kappa_2\tanh\kappa_1\cosh^{-1}\kappa_1+\mathcal{O}(\kappa_2^2)\,,
    \end{split}
\end{equation}
which coincides with the JT brane \eqref{jt-locate-1} upon the identification ,
\begin{equation}\label{identification}
    \kappa_1= \rho_0,\qquad
    \kappa_2={\rho_0\bar\phi_r}=6L\mu.
\end{equation}
The explicit derivation of the above results is given .

Again we assume entanglement islands can emerge on the left hand side $x<0$, and calculate the entanglement entropy for intervals on the right hand side $x>0$ via both the island formula and the RT formula. Then compare the results with those in AdS/BCFT with fluctuating brane. Let us consider the interval $A=[0,L]$ and assume the island region $\is (A)=[-a,0)$, then the island formula gives
\begin{equation}\label{ee-wcft}
\begin{split}
    S([0,L])&=\min_a\left\{\frac{c}{6}\log\frac{(L+a)^2}{2a\epsilon}+\frac{c}{6}\kappa_1+\frac{c}{6}\frac{\kappa_2}{a}\right\}\\
    &\approx\frac{c}{6}\log\frac{2L}{\epsilon}+\frac{c}{6}\kappa_1+\frac{c}{6}\frac{\kappa_2}{L},
    \end{split}
\end{equation}
which aligns with \eqref{Omega2Sa} upon the identifications \eqref{identification}. 
The above formula is minimized at
\begin{equation}\label{ais-w}
    x\approx -a_{\rm bdy}=-L-12L\mu,
\end{equation}
which also coincides with \eqref{is-a-2}.

Holographically, $S_A$ is given by the RT surface connecting the two boundary points at $x=-a_{\rm bdy},L$, which is cut off at the cutoff sphere centered at $x=-a_{\rm bdy}$. 
Note that the point $x=-a_{\rm bdy}$ locates on the asymptotic boundary, i.e. $y=-a_{\rm bdy},\rho\to\infty$, while the cutoff point $y=-a_{\rm brane}$ is on the cutoff brane.
Since $a_{\rm bdy}\neq L$, the center of the RT curve is no longer the origin, rather it is settled at
\begin{equation}\label{x0}
    x=-x_0\approx\frac{-a_{\rm bdy}+L}{2}=-6L\mu.
\end{equation}
We now need to check two features for the RT curve to match it with the RT surface anchored on the KR brane in AdS/BCFT. Firstly, the cutoff point on the RT curve should lie on the cutoff brane. 
Secondly, the RT curve should be normal to the cutoff brane at the cutoff point. 
If the first feature holds, then the tangent line of the cutoff brane at the cutoff point should pass the center of this RT curve, as shown in the right figure of Fig. \ref{fig:WCFT2}. According to \eqref{tangentbrane}, the intersection point of this tangent line and the asymptotic boundary is given by 
\begin{align}
x=-\frac{\kappa_2 a_{\rm bdy}}{\kappa_2+a_{\rm bdy}}\approx -6 L\mu\,,
\end{align}
which, as expected, coincides with the center of the RT circle up to order $\mathcal{O}(\mu)$, hence the first feature is confirmed. 
Moreover, since the RT curve is normal to the cutoff sphere and the cutoff brane is tangent to the cutoff sphere, the RT curve should be normal to the cutoff brane, thus the second feature is confirmed. 
Furthermore, the $y$ coordinate of the cutoff point is given by
\begin{equation}\label{ain-w}
   y= -a_{\rm brane}\approx -L-6L(\tanh\rho_0+1)\mu\neq -a_{\rm bdy},
\end{equation}
which coincides with the point \eqref{RTsaddle} at which the RT surface anchors on the KR brane in AdS/BCFT.

The fact that $a_{\rm bdy}\neq a_{\rm brane}$ is well understood in the Weyl transformed CFT$_2$ set-up. 
The naive reason is that, due to the fluctuation of the brane, the RT circle of the interval $[0,L]$ is no longer centered as the origin, which is obvious in Fig.\ref{fig:WCFT2}. 
Nevertheless, this inequality raises up an important physical confusion in the set-up of JT gravity on the brane coupled to a bath CFT$_2$. 
In this set-up, since the gravity part of the $2d$ effective theory is on the KR brane, the additional twist operator due to the island formula should be inserted at $y=-a_{\rm brane}$, which is also the point where the RT curve anchors on the brane. This indicates $a_{\rm bdy}=a_{\rm brane}$, which is inconsistent with our observation. 
As we have shown, this insertion will not reproduce the result from the RT formula in doubly holography. Our discussion indicates that the twist operator should be inserted at $y=-a_{\rm bdy}$. 
This confusion can be fully understood in the holographic Weyl transformed set-up, where the dual field theory is always on the asymptotic boundary rather than the cutoff brane. This leads us to a sharp conjecture that, all the AdS/BCFT set-ups are essentially a holographic Weyl transformed CFT$_2$.

\section{Entanglement wedge cross-sections with fluctuating brane}\label{sec:ew-jt}
Here we classify all possible phases for the entanglement wedge of $A\cup B$ with a fluctuating KR brane, and calculate the area of the corresponding EWCS. We set $A=[b_1,b_2]$ and $B=[b_3,b_4]$ with $0\leq b_1<b_2\leq b_3<b_4$. 
Adjacency between regions $A$ and $B$ happens when $b_2=b_3$. 
We present illustrations of all classifications for EWCS in Figs.\ \ref{fig:ew-A1} and \ref{fig:ew-A2}. Compared with the configuration with non-fluctuating brane \cite{Basu:2023wmv}, computing the area of EWCS is more complicated in the present scenario. Fortunately, as we will show below, perturbing the results in the non-fluctuating configurations allows us to obtain the area of EWCS up to the first order easily.

\subsection{No-island phase}
\begin{figure}
    \centering
    \includegraphics[width=0.9\textwidth]{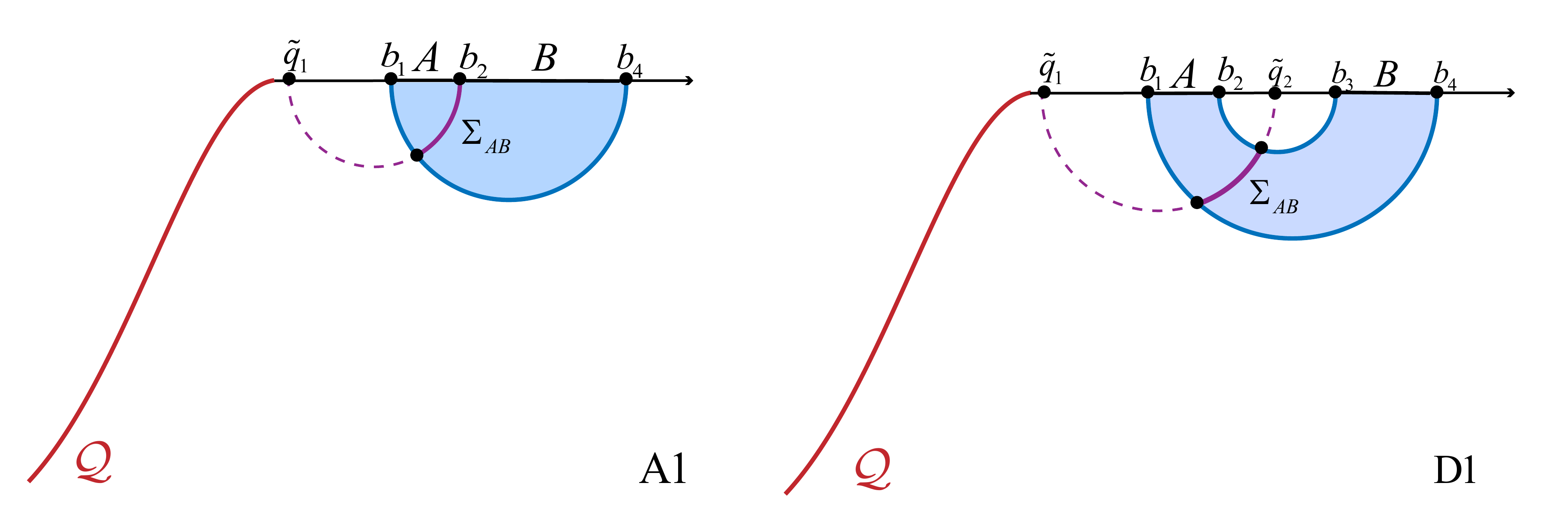}
    \caption{EWCS for phases A1 and D1
    }
    \label{fig:ew-A1}
\end{figure}
We first consider the simplest cases, where $AB$ admits no island, hence the EWCS coincides with those in AdS$_3$/CFT$_2$. We denote such no-island phases as phase A1 and B1 (see in Fig.\ \ref{fig:ew-A1}), where $A$ and $B$ are adjacent and disjoint respectively. The corresponding area of the EWCSs are those given in \cite{Takayanagi:2017knl}
\begin{equation}\label{ew-A1D1}
\begin{split}
  &  \textbf{Phase-A1 :} \frac{\operatorname{Area}\left[\Sigma_{A B}\right]}{4 G_N}=\frac{c}{6} \log \frac{2\left(b_2-b_1\right)\left(b_4-b_2\right)}{\epsilon\left(b_4-b_1\right)},\\
  & \textbf{Phase-D1 :} \frac{\operatorname{Area}\left[\Sigma_{A B}\right]}{4 G_N}=\frac{c}{6} \log \frac{1+\sqrt{w}}{1-\sqrt{w}},\quad w=\frac{\left(b_4-b_3\right)\left(b_2-b_1\right)}{\left(b_3-b_1\right)\left(b_4-b_2\right)}.
  \end{split}
\end{equation}

\subsection{Island phases}
\begin{figure}
    \centering
    \includegraphics[width=0.95\textwidth]{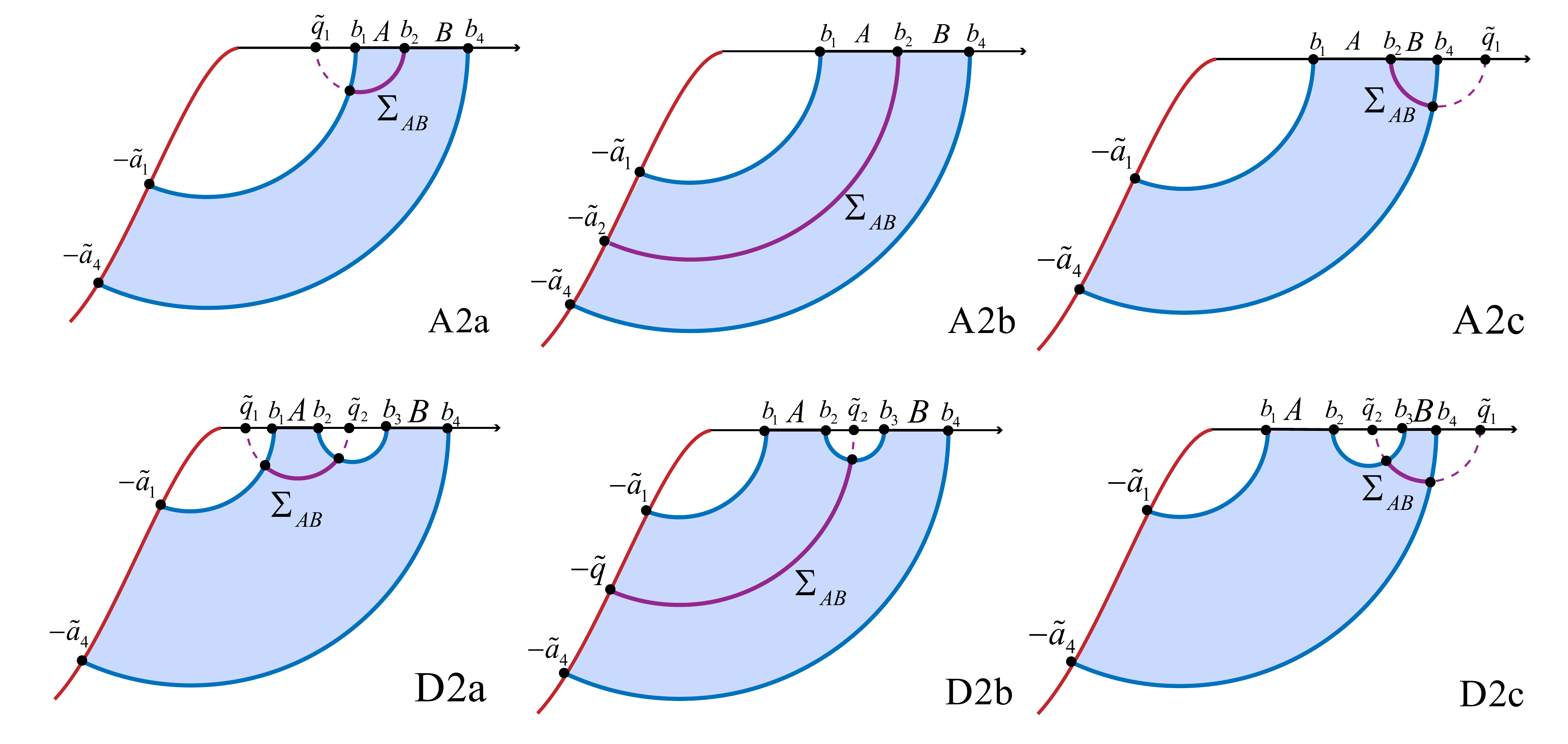}
    \caption{EWCS for phases A2 and D2
    }
    \label{fig:ew-A2}
\end{figure}

Now we consider the cases where $AB$ admits an island. When $A$ and $B$ are adjacent, we denote the phases as A2, otherwise we denote them as D2.
According to the position where the EWCS anchors in the bulk, phases A2 and D2 can further be decomposed into the following three classes:
\begin{itemize}
    \item Phases A2a and D2a:
    the EWCS is anchored on the RT($b_1$),
    \item Phases A2b and D2b:
    the EWCS is anchored on the brane,
     \item Phases A2c and D2c:
      the EWCS is anchored on the RT($b_4$),
\end{itemize}
where RT($b_i$) denotes the RT surface that emanates from the boundary point $x=b_i$ and anchors on the brane (see Fig.\ \ref{fig:ew-A2}).

As was pointed out in Sec.\ \ref{sec:wtcft}, compared with the non-fluctuating configurations, the center of the RT($b_i$) circle is no longer the origin $x=0$. 
Rather, it is settled at \eqref{x0}
\begin{align}
x=-x_0=-6 b_i \mu=-\rho_0 \bar{\phi}_{r}\,.
\end{align}
It is important to note that, $x_0$ is independent of $b_i$, hence the centers of all RT$(b_i)$ circles are the same. For convenience, later we will expand our results with respect to $x_0$.

\textbf{In phase A2a}, the EWCS emanates from the boundary point $x=b_2$ and anchors on RT($b_1$) vertically. Since the center of RT($b_i$) settles at $x=-x_0$, the whole configuration differs from the non-fluctuating one by a simple translation $x\to x+x_0$.  
Therefore, the area of the EWCS can be easily obtained from the results in the non-fluctuating configuration, with the replacement $b_1\to b_1+x_0$ and $b_2\to b_2+x_0$,
\begin{align}\label{a2aew}
   \frac{\text{Area}(\Sigma_{AB})}{4G_N}=&\frac{c}{6}\log\frac{b_2^2-b_1^2}{b_1\epsilon}\Big{|}_{b_1\to b_1+x_0,~b_2\to b_2+x_0}
   \cr
   =&\frac{c}{6}\log\frac{(b_2+x_0)^2-(b_1+x_0)^2}{(b_1+x_0)\epsilon}
   \cr
   \approx& 
\frac{c}{6}\log\frac{b_2^2-b_1^2}{b_1\epsilon}-\frac{c}{6}\frac{b_2-b_1}{b_1+b_2}\frac{x_0}{b_1}.
\end{align}
One can extend the EWCS to the asymptotic boundary and find intersection point is given by
\begin{equation}\label{A2a-tq1}
    \tilde q_1=\frac{(b_1+x_0)^2}{b_2+x_0}-x_0
    \approx
    \frac{b_1^2}{b_2}-\frac{(b_1-b_2)^2}{b_2^2}x_0.
\end{equation}
We will see that, this intersection point coincides with the balance point, which solves the balance requirements. 

\textbf{In phase A2b}, the EWCS is just the RT($b_2$) circle, whose area is given by  \eqref{doublehsa} with the replacement $L\to b_2$,
\begin{equation}\label{ew-a2b}
\begin{aligned}
\frac{\text{Area}(\Sigma_{AB})}{4G_N}& =
\frac{c}{6}\log\frac{2b_2}{\epsilon}+\frac{c}{6} \rho_0+\frac{c}{6}\frac{x_0}{b_2}.
\end{aligned}
\end{equation}

\textbf{In phase A2c}, the EWCS in phase A2c anchors on the RT($b_4$) circle, and its area can be calculated in similar way, which is just the result in the non-fluctuating configurations with a translation,
\begin{equation}\label{ew-a2c}
\begin{split}
   \frac{\text{Area}(\Sigma_{AB})}{4G_N}
   &=\frac{c}{6}\log\frac{b_4^2-b_2^2}{b_4\epsilon}\Big{|}_{b_2\to b_2+x_0,~b_4\to b_4+x_0}
   \cr
   &\approx
   \frac{c}{6}\log\frac{b_4^2-b_2^2}{b_4\epsilon}-\frac{c}{6}\frac{b_4-b_2}{b_4+b_2}\frac{x_0}{b_4}.
\end{split}\end{equation}
Extending the EWCS to the asymptotic boundary, we get
\begin{equation}\label{A2c-tq1}
    \tilde q_1
    \approx
    \frac{b_4^2}{b_2}-\frac{(b_4-b_2)^2}{b_2^2}x_0.
\end{equation}

\textbf{In phase D2a}, the EWCS anchors on the RT($b_1$) circle and the RT surface for the interval $[b_2,b_3]$. 
The area of the EWCS is given by
\begin{equation}\label{ew-d2a}
\begin{split}
    \frac{\text{Area}(\Sigma_{AB})}{4G_N}
=&\frac{c}{6} \log \frac{b_2b_3-b_1^2+\sqrt{(b_3^2-b_1^2)(b_2^2-b_1^2)}}{b_1\left(b_3-b_2\right)}\Big{|}_{b_1\to b_1+x_0,~b_2\to b_2+x_0}\\
=&\frac{c}{6} \log\left( \frac{b_2 b_3-b_1^2+\sqrt{\left(b_2^2-b_1^2\right)\left(b_3^2-b_1^2\right)}}{b_1\left(b_3-b_2\right)}\right)
-\frac{c}{6}\frac{\sqrt{\left(b_2^2-b_1^2\right)\left(b_3^2-b_1^2\right)}}{b_1(b_1+b_2)(b_1+b_3)}x_0.
\end{split}
\end{equation}
Extend the EWCS to the asymptotic boundary, and we get
\begin{equation}\label{tq1}
    \begin{split}
        \tilde q_1\approx& \frac{b_1^2+b_2b_3-\sqrt{(b_2^2-b_1^2)(b_3^2-b_1^2)}}{b_2+b_3}
        -\left(
        \frac{2(b_1^2+b_2b_3-\sqrt{(b_2^2-b_1^2)(b_3^2-b_1^2)})}{(b_2+b_3)^2}\right.\\
        &\left.+\frac{-2b_1(b_1+b_2)(b_1+b_3)+\sqrt{(b_2^2-b_1^2)(b_3^2-b_1^2)}(2b_1+b_2+b_3)}{(b_2+b_3)(b_1+b_2)(b_1+b_3)}
        \right)x_0,
    \end{split}
\end{equation}
\begin{equation}\label{tq2}
    \begin{split}
        \tilde q_2\approx& \frac{b_1^2+b_2b_3+\sqrt{(b_2^2-b_1^2)(b_3^2-b_1^2)}}{b_2+b_3}
        -\left(
        \frac{2(b_1^2+b_2b_3+\sqrt{(b_2^2-b_1^2)(b_3^2-b_1^2)})}{(b_2+b_3)^2}\right.\\
        &\left.+\frac{-2b_1(b_1+b_2)(b_1+b_3)-\sqrt{(b_2^2-b_1^2)(b_3^2-b_1^2)}(2b_1+b_2+b_3)}{(b_2+b_3)(b_1+b_2)(b_1+b_3)}
        \right)x_0.
    \end{split}
\end{equation}

\textbf{In phase D2b}, the EWCS anchors on the brane and the RT surface for the interval $[b_2,b_3]$. The derivation for the area of EWCS gets tricky.
We leave it in Appendix.\ \ref{B}.
The results are as follows.
The area of EWCS is
\begin{equation}\label{ew-d2b}
    \begin{split}
  \frac{\text{Area}(\Sigma_{AB})}{4G_N}=  \frac{c}{6}\log\frac{\sqrt{b_3}+\sqrt{b_2}}{\sqrt{b_3}-\sqrt{b_2}}+\frac{c}{6} \rho_0+\frac{c}{6}\frac{x_0}{\sqrt{b_2b_3}}.
    \end{split}
\end{equation}
And the intersection point with the asymptotic boundary of AdS is 
\begin{equation}\label{inter-d2b}
    \tilde q_2=\sqrt{b_2b_3}+3(\sqrt{b_3}-\sqrt{b_2})^2\frac{x_0}{6\sqrt{b_2b_3}}.
\end{equation}

\textbf{In phase D2c}, the EWCS anchors on RT$(b_4)$.
Similar to phase D2c, the area of the EWCS for phase D2c can be easily obtained from the non-fluctuating case with a translation,
\begin{equation}\label{ew-d2c}
\begin{split}
    \frac{\text{Area}(\Sigma_{AB})}{4G_N}
=&\frac{c}{6} \log\left( \frac{b_2 b_3-b_4^2+\sqrt{\left(b_2^2-b_4^2\right)\left(b_3^2-b_4^2\right)}}{b_4\left(b_3-b_2\right)}\right)\Big{|}_{b_4\to b_4+x_0,~b_2\to b_2+x_0}\\
=&\frac{c}{6} \log\left( \frac{b_2 b_3-b_4^2+\sqrt{\left(b_2^2-b_4^2\right)\left(b_3^2-b_4^2\right)}}{b_4\left(b_3-b_2\right)}\right)-\frac{c}{6}\frac{\sqrt{\left(b_2^2-b_4^2\right)\left(b_3^2-b_4^2\right)}}{b_4(b_4+b_2)(b_4+b_3)}x_0.
\end{split}
\end{equation}
Two intersection points are
\begin{equation}\label{D2c-tq1}
    \begin{split}
        \tilde q_1&\approx \frac{b_4^2+b_2b_3+\sqrt{(b_2^2-b_4^2)(b_3^2-b_4^2)}}{b_2+b_3}
        -\left(
        \frac{2(b_4^2+b_2b_3+\sqrt{(b_2^2-b_4^2)(b_3^2-b_4^2)})}{(b_2+b_3)^2}\right.\\
        &\left.+\frac{-2b_4(b_4+b_2)(b_4+b_3)-\sqrt{(b_2^2-b_4^2)(b_3^2-b_4^2)}(2b_4+b_2+b_3)}{(b_2+b_3)(b_4+b_2)(b_4+b_3)}
        \right)x_0,
    \end{split}
\end{equation}
\begin{equation}\label{D2c-tq2}
    \begin{split}
        \tilde q_2&\approx \frac{b_4^2+b_2b_3-\sqrt{(b_2^2-b_4^2)(b_3^2-b_4^2)}}{b_2+b_3}
        -\left(
        \frac{2(b_4^2+b_2b_3-\sqrt{(b_2^2-b_4^2)(b_3^2-b_4^2)})}{(b_2+b_3)^2}\right.\\
        &\left.+\frac{-2b_4(b_4+b_2)(b_4+b_3)+\sqrt{(b_2^2-b_4^2)(b_3^2-b_4^2)}(2b_4+b_2+b_3)}{(b_2+b_3)(b_4+b_2)(b_4+b_3)}
        \right)x_0.
    \end{split}
\end{equation}

\section{BPE in Weyl transformed CFT}\label{sec:bpe-jt}
In this section, we study the BPE in the AdS/BCFT set-up with fluctuating brane, and match it with the area of the EWCS.
The computation of the BPE in the fluctuating brane configurations is similar to that in non-fluctuating configurations \cite{Basu:2023wmv}, and our foundation can be summarized into the following three proposals:
\begin{enumerate}
\item As in the no-island phases, the PEE structure of the island phase is also described by the two-point PEEs $\mathcal{I}(x,y)$.

\item When computing the PEE between any two spacelike separated regions, we should also properly take into account the contributions from their island regions as well as from the ownerless islands.

\item The two-point functions between the endpoints of any interval $[a,b]$ in the Weyl transformed CFT are well defined and given by \eqref{twopointf}. Furthermore, this two-point function represents the PEE following the \textit{basic proposal} \eqref{basicp}. 
\end{enumerate}

Compared with the non-fluctuating configurations \cite{Basu:2023wmv}, the position where the EWCS anchors on the brane and the position of the balance point on the asymptotic boundary differ from each other, which is akin to the observation that $a_{\rm bdy}\neq a_{\rm brane}$ in our discussion on the entanglement entropy.

The simplest cases are the phases A1 and D1, where $AB$ admits no island and the correlations between $A$ and $B$ receive no contribution from the islands. In these cases, the PEE, BPE and EWCS coincide with their counterpart in AdS$_3$/CFT$_2$, see Fig.\ref{fig:ew-A1}. In the following, we focus on the cases where $AB$ admits a island. 

\subsection{Adjacent $AB$ with island}
\begin{figure}
    \centering
    \includegraphics[width=0.8\textwidth]{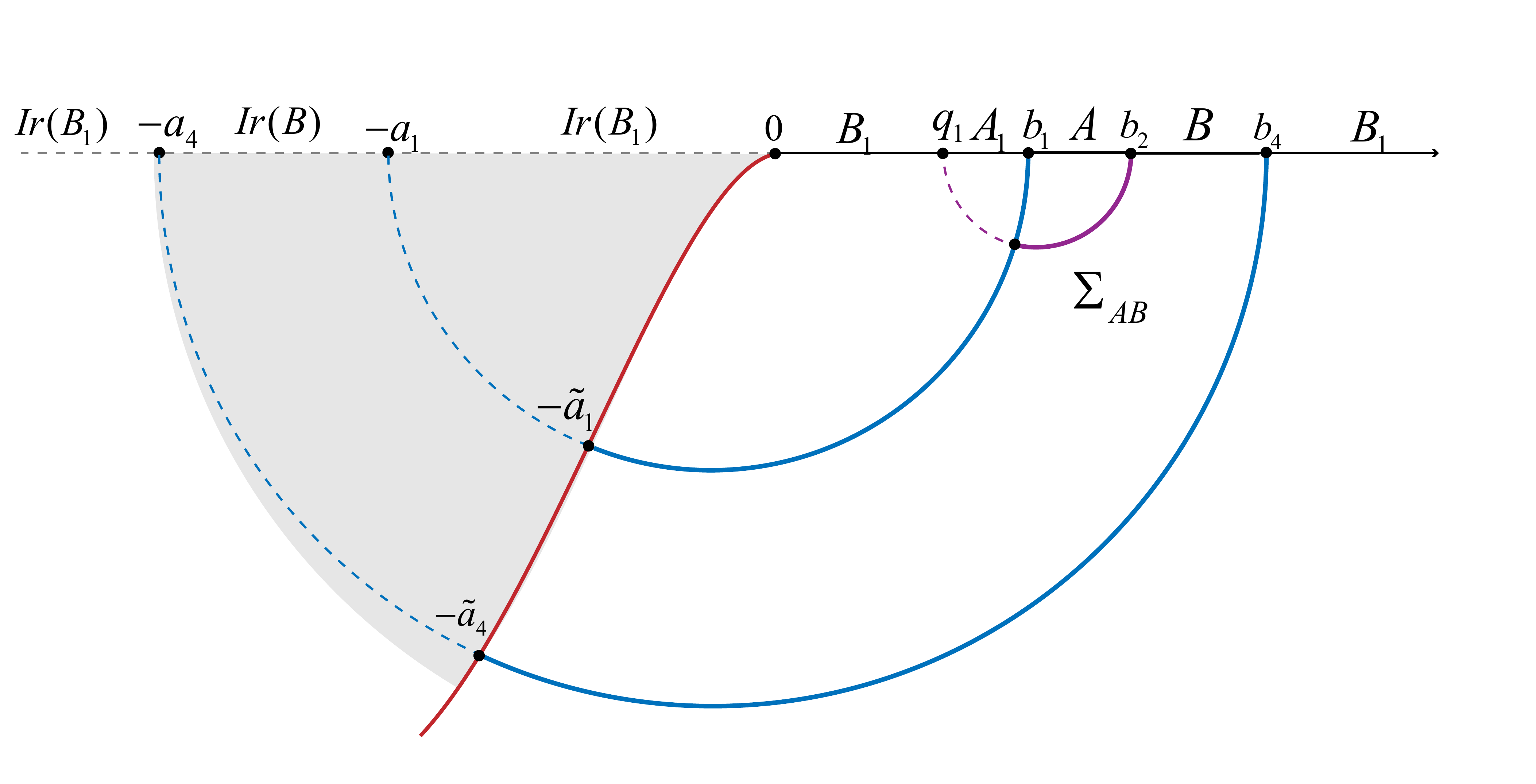}
    \caption{Phase A2a.
    }
    \label{fig:jt-A2a}
\end{figure}

\begin{figure}
    \centering
    \includegraphics[width=0.8\textwidth]{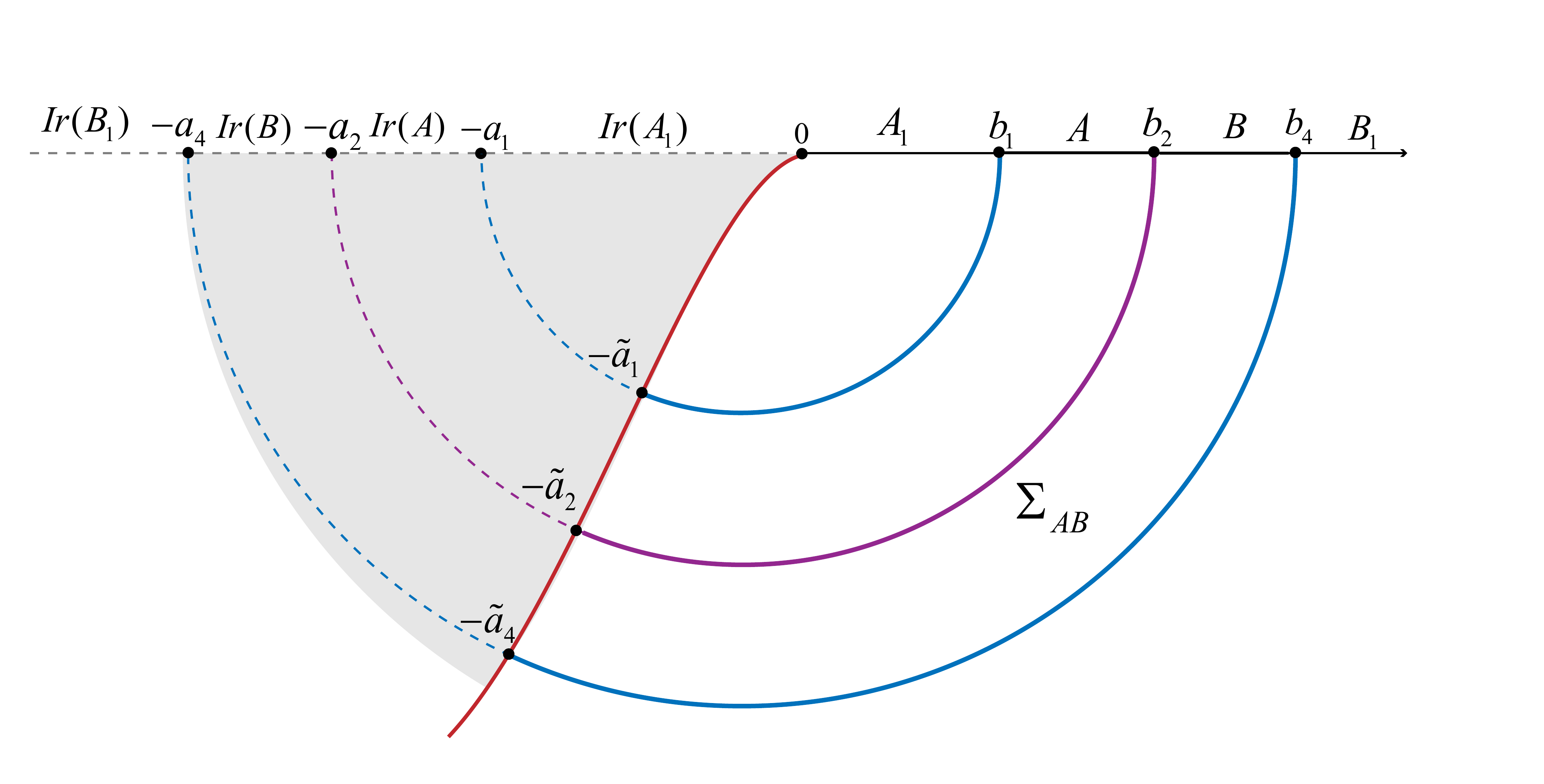}
    \caption{Phase A2b
    }
    \label{fig:jt-A2b}
\end{figure}

\begin{figure}
    \centering
    \includegraphics[width=0.8\textwidth]{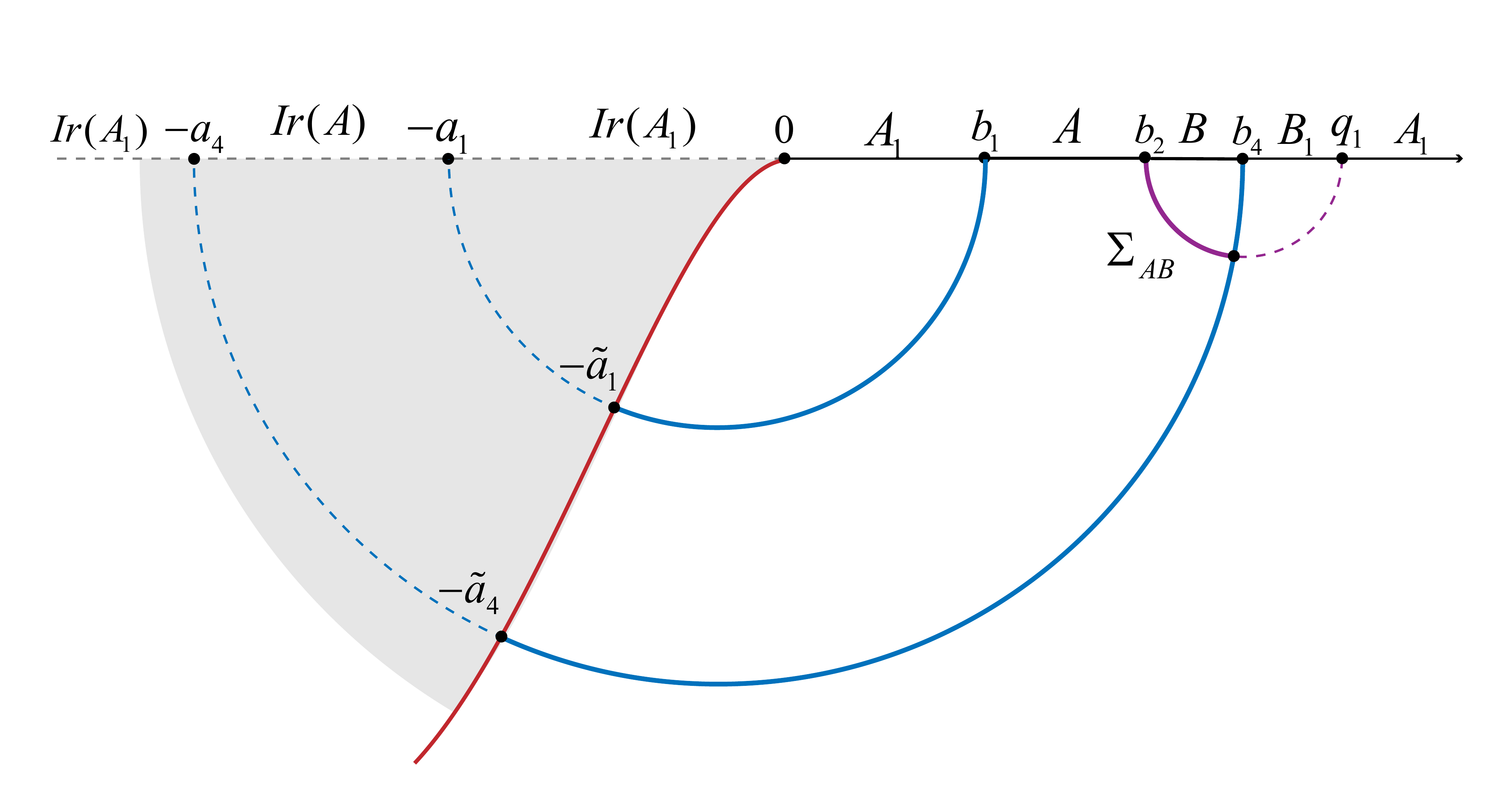}
    \caption{Phase A2c
    }
    \label{fig:jt-A2c}
\end{figure}

Let us first consider the adjacent case, where the island region $\is (AB)$ in the Weyl transformed CFT is given by $\text{Is}(AB)=[-a_4,-a_1]$ with $a_i\approx b_i+2x_0\,,~i=1,4$. And the cross-section has three different saddle points, which are shown in Figs. \ref{fig:jt-A2a}, \ref{fig:jt-A2b} and \ref{fig:jt-A2c} respectively. The assignment for the ownerless island region also has three possibilities:
\begin{enumerate}
    \item $\operatorname{Ir}(A)=\emptyset, \quad \operatorname{Ir}(B)=[-a_4,-a_1]$.
    \item $\operatorname{Ir}(A)=[-q,-a_1], \quad \operatorname{Ir}(B)=[-a_4,-q]$ with $a_1<q<a_4$.
    \item $\operatorname{Ir}(A)=[-a_4,-a_1], \quad \operatorname{Ir}(B)=\emptyset$.
\end{enumerate}
Following the discussions in \cite{Basu:2023wmv}, for the above assignments 1, 2, and 3, we can solve the balance requirements and get a corresponding BPE, which exactly matches the three saddles of the cross-section shown in Figs. \ref{fig:jt-A2a}, \ref{fig:jt-A2b} and \ref{fig:jt-A2c}. 
We ought to select the assignment that yields the minimal BPE, corresponding to the area of minimal cross-section, i.e. the EWCS.

\textbf{In phase A2a}, the EWCS anchors on RT($b_1)$ and we should consider the assignment $\operatorname{Ir}(A)=\emptyset$, $\operatorname{Ir}(B)=[-a_4,-a_1]$. Let us assume 
\begin{equation}
 \quad \text{Ir}(A_1)=\emptyset,\quad
    \text{Ir}(B_1)=(-\infty,-a_4]\cup[-a_1,0].
\end{equation}
In this case the partition point satisfies $0<q_1<b_1$, see Fig.\ \ref{fig:jt-A2a}.

Now we solve the balanced requirements. Using the generalized ALC formula \eqref{galc1}, we get the two relevant PEEs 
\begin{equation}
\begin{aligned}
\mathcal{I}\left(A, B \operatorname{Ir}(B)  B_1 \operatorname{Ir}\left(B_1\right)\right) & =\frac{1}{2}\left[\tilde{S}_{A A_1}+\tilde{S}_A-\tilde{S}_{A_1}\right] \\
& =\frac{1}{2}\left(\tilde{S}_{\left[q_1, b_2\right]}+\tilde{S}_{\left[b_1, b_2\right]}-\tilde{S}_{\left[q_1, b_1\right]}\right)\\
&=
\frac{c}{6}\log\frac{(b_2-q_1)(b_2-b_1)}{(b_1-q_1)\epsilon},\\ \cr
\mathcal{I}\left(B \operatorname{Ir}(B), A A_1\right) & =\frac{1}{2}\left[\tilde{S}_{B \operatorname{Ir}(B) B_1 \operatorname{Ir}\left(B_1\right)}+\tilde{S}_{B \operatorname{Ir}(B)}-\tilde{S}_{B_1 \operatorname{Ir}\left(B_1\right)}\right] \\
& =\frac{1}{2}\left(\tilde{S}_{\left[q_1, b_2\right]}+\tilde{S}_{\left[-b_4,-b_1\right] \cup\left[b_2, b_4\right]}-\tilde{S}_{\left[-\infty,-b_4\right] \cup\left[-b_1, q_1\right] \cup\left[b_4, \infty\right]}\right) \\
& =\frac{c}{6} \log \frac{\left(b_2-q_1\right)\left(b_2+a_1\right)}{\left(q_1+a_1\right) \epsilon} .
\end{aligned}
\end{equation}
Then solving the balance condition $\mathcal{I}\left(A, B \operatorname{Ir}(B) \cup B_1 \operatorname{Ir}\left(B_1\right)\right) =\mathcal{I}\left(B \operatorname{Ir}(B), A A_1\right) $,
we obtain the balance point
\begin{equation}\label{bp-a2a}
    q_1\approx  \frac{b_1^2}{b_2}-\frac{(b_1-b_2)^2}{b_2^2}x_0.
\end{equation}
Furthermore, we can get the following BPE,
\begin{equation}\label{bpe-a2a}
\begin{split}
    \text{BPE}(A:B)=&
    \frac{c}{6}\log\frac{(b_2-q_1)}{\epsilon}\frac{(b_2-b_1)}{(b_1-q_1)}
    \approx
    \frac{c}{6}\log\frac{b_2^2-b_1^2}{b_1\epsilon}-\frac{c}{6}\frac{b_2-b_1}{b_1+b_2}\frac{x_0}{b_1},
    \end{split}
\end{equation}
which matches the area of EWCS \eqref{a2aew} exactly. 

One can also assume that $\text{Ir}(A_1)\neq \emptyset$. Nevertheless, similar to the non-fluctuating configurations studied in \cite{Basu:2023wmv}, the solution to the balance requirements does not exist under such an assumption.

\textbf{In Phase A2b}, the EWCS anchors on the brane and we should consider $\operatorname{Ir}(A)=[-q,-a_1]$, and $ \operatorname{Ir}(B)=[-a_4,-q]$.
Since $A$ and $B$ in this phase are symmetric in the sense that they both receive island contributions, it is natural to assume both $A_1$ and $B_1$ admit island contributions $\text{Ir}(A_1)$ and $\text{Ir}(B_1)$.
In particular, we choose
\begin{equation}
    A_1=[0,a_1],\quad 
    B_1=[a_4,+\infty),
\end{equation}
and
\begin{equation}
    \text{Ir}(A_1)=[-a_1,0],\quad
    \text{Ir}(B_1)=(-\infty,-a_4].
\end{equation}
Then using the generalized ALC formula \eqref{galc1}, we get 
\begin{equation}
\begin{aligned}
\mathcal{I}\left(A \operatorname{Ir}(A), B \operatorname{Ir}(B) B_1 \operatorname{Ir}\left(B_1\right)\right) & =\frac{1}{2}\left[\tilde{S}_{A \operatorname{Ir}(A) A_1 \operatorname{Ir}\left(A_1\right)}+\tilde{S}_{A \operatorname{Ir}(A)}-\tilde{S}_{A_1 \operatorname{Ir}\left(A_1\right)}\right] \\
& =\frac{1}{2}\left[\tilde{S}_{\left[-q, b_2\right]}+\tilde{S}_{\left[-q,-b_1\right] \cup\left[b_1, b_2\right]}-\tilde{S}_{\left[-b_1, b_1\right]}\right] \\
& =\tilde{S}_{\left[-q, b_2\right]},
\end{aligned}
\end{equation}
\begin{equation}
\begin{aligned}
\mathcal{I}\left(B \operatorname{Ir}(B), A \operatorname{Ir}(A) A_1 \operatorname{Ir}\left(A_1\right)\right) & =\frac{1}{2}\left[\tilde{S}_{B \operatorname{Ir}(B) B_1 \operatorname{Ir}\left(B_1\right)}+\tilde{S}_{B \operatorname{Ir}(B)}-\tilde{S}_{B_1 \operatorname{Ir}\left(B_1\right)}\right] \\
& =\frac{1}{2}\left[\tilde{S}_{\left[-q, b_2\right]}+\tilde{S}_{\left[-b_4,-q\right] \cup\left[b_2, b_4\right]}-\tilde{S}_{\left[-b_4, b_4\right]}\right] \\
& =\tilde{S}_{\left[-q, b_2\right]} ,
\end{aligned}
\end{equation}
with
\begin{equation}\label{S_a2b}
    \tilde{S}_{\left[-q, b_2\right]}
    =\frac{c}{6}\log\frac{(q+b_2)^2}{2 q\epsilon}+\frac{c}{6}\rho_0+\frac{c}{6}\frac{x_0}{q}.
\end{equation}
Thus the balanced requirement is automatically satisfied regardless of the choice of $q\in [-a_4,-a_1]$.
However, different $q$ will give different BPEs, and according to the minimal requirement, the minimal one is taken at
\begin{equation}\label{eq:q_A2b}
    \frac{\D}{\D q}\mathcal{I}\left(A \operatorname{Ir}(A), B \operatorname{Ir}(B) B_1 \operatorname{Ir}\left(B_1\right)\right)=0\quad\Rightarrow \quad q=b_2+2x_0.
\end{equation}
Inserting \eqref{eq:q_A2b} into \eqref{S_a2b}, we get the BPE 
\begin{equation}
\text{BPE}(A:B)=\frac{c}{6} \rho_0+\frac{c}{6}\log\frac{2b_2}{\epsilon}+\frac{c}{6}\frac{x_0}{b_2},
\end{equation}
which is exactly the area of EWCS \eqref{ew-a2b}.

\textbf{In Phase A2c}, the EWCS anchors on RT$(b_4)$ and we should consider  $\operatorname{Ir}(A)=[-a_4,-a_1]$ and $\operatorname{Ir}(B)=\emptyset$.
Following the similar argument as phase A2a, we arrive at 
\begin{equation}
    q_1\approx  \frac{b_4^2}{b_2}-\frac{(b_4-b_2)^2}{b_2^2}x_0,
\end{equation}
\begin{equation}
\begin{split}
    \text{BPE}(A:B)
    \approx& 
     \frac{c}{6}\log\frac{b_4^2-b_2^2}{b_4\epsilon}-\frac{c}{6}\frac{b_4-b_2}{b_2+b_4}\frac{x_0}{b_4},
    \end{split}
\end{equation}
which coincides with the area of the EWCS \eqref{ew-a2c}.

\subsection*{Minimizing BPE for phase A2}
The last step of the generalized ALC proposal is to choose the minimal BPE.
Now let us compare the different BPEs in phase A2.

The critical point between phase A2a and A2b is given by
\begin{equation}
    f(b_1,b_2)\equiv\rho_0+\log\frac{2b_1b_2}{(b_2^2-b_1^2)\epsilon}+\frac{(b_1^2+b_2^2)x_0}{b_1b_2(b_1+b_2)}=0.
\end{equation}
Phase A2a has the smaller BPE when $f(b_4,b_2)>0$.
Note that we assume $A_1A$ admits no island in phase A2a.
Now we show that this assumption is consistent with the minimal condition $f(b_4,b_2)>0$ for phase A2a. 
Let us compare the island and non-island saddles of the entanglement entropy for  $A_1A$
\begin{equation}
\begin{aligned}
&S_{\mathrm{island}}\left(A_1 A \right)-S_{\mathrm{non-island}}\left(A_1 A \right)\\
= & \frac{c}{3}\rho_0+ \frac{c}{6}\log\frac{4q_1b_2}{\epsilon^2}+\frac{c}{6}\left(\frac{1}{b_2}+\frac{1}{q_1}\right)x_0-\frac{c}{6}\log\frac{(b_2-q_1)^2}{\epsilon^2}\\
=&\frac{c}{3}f(b_1,b_2)>0,
\end{aligned}
\end{equation}
where we have used the balance point \eqref{bp-a2a}.
Thus when phase A2a takes the smaller BPE, $A_1A$ admits no island.

Similarly, the critical point between phase A2c and A2b is given by
\begin{equation}
    f(b_4,b_2)\equiv\rho_0+\log\frac{2b_4b_2}{(b_4^2-b_2^2)\epsilon}+\frac{(b_2^2+b_4^2)x_0}{b_2b_4(b_2+b_4)}=0.
\end{equation}
And phase A2c has the smaller BPE when $f(b_4,b_2)>0$.
Follow the similar argument, one could show that $BB_1$ admits no island in phase A2c under $f(b_4,b_2)>0$.

\subsection{Disjoint $AB$ with island}

\begin{figure}
    \centering
    \includegraphics[width=0.8\textwidth]{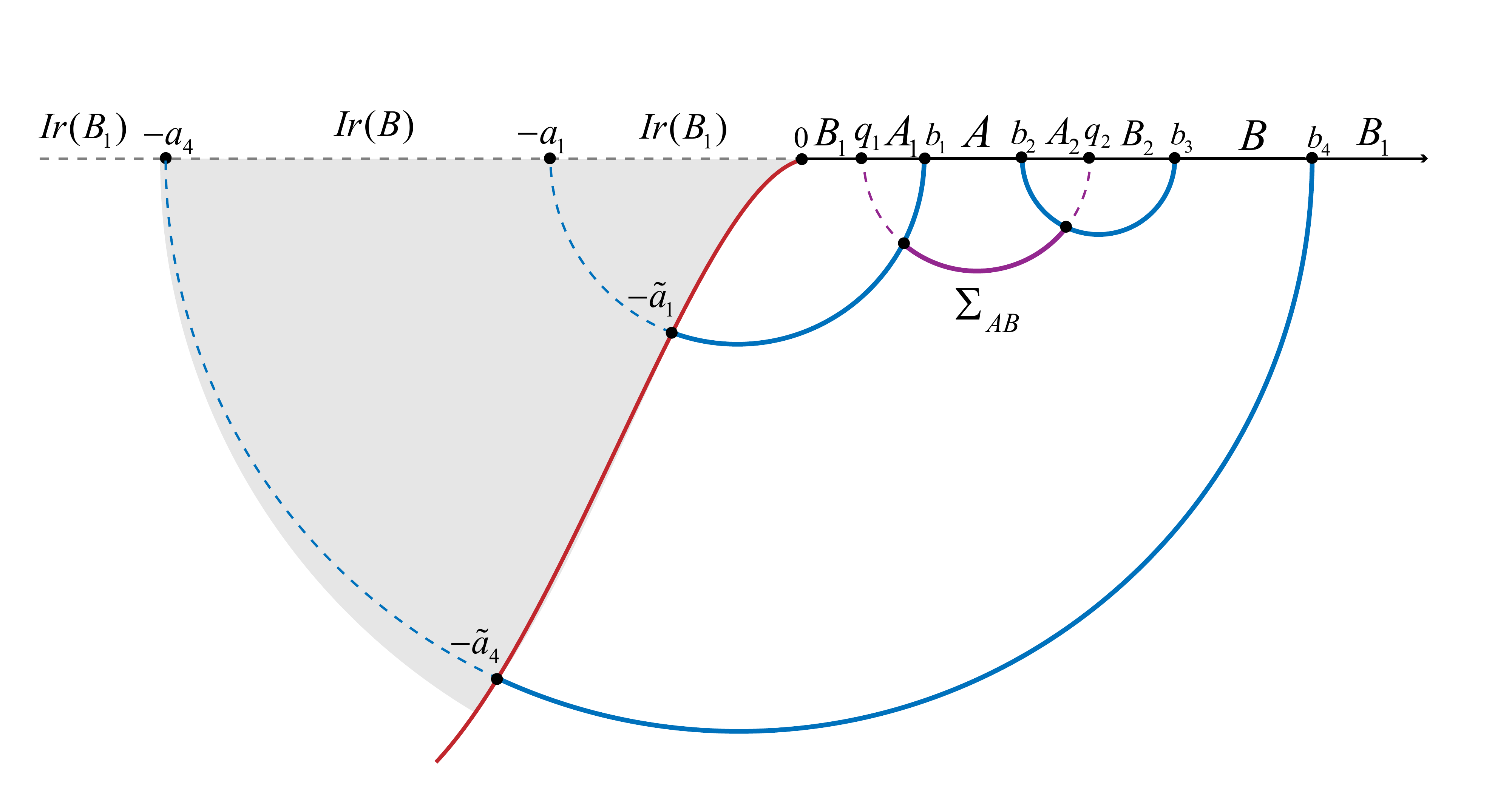}
    \caption{Phase D2a
    }
    \label{fig:jt-D2a}
\end{figure}

\begin{figure}
    \centering
    \includegraphics[width=0.8\textwidth]{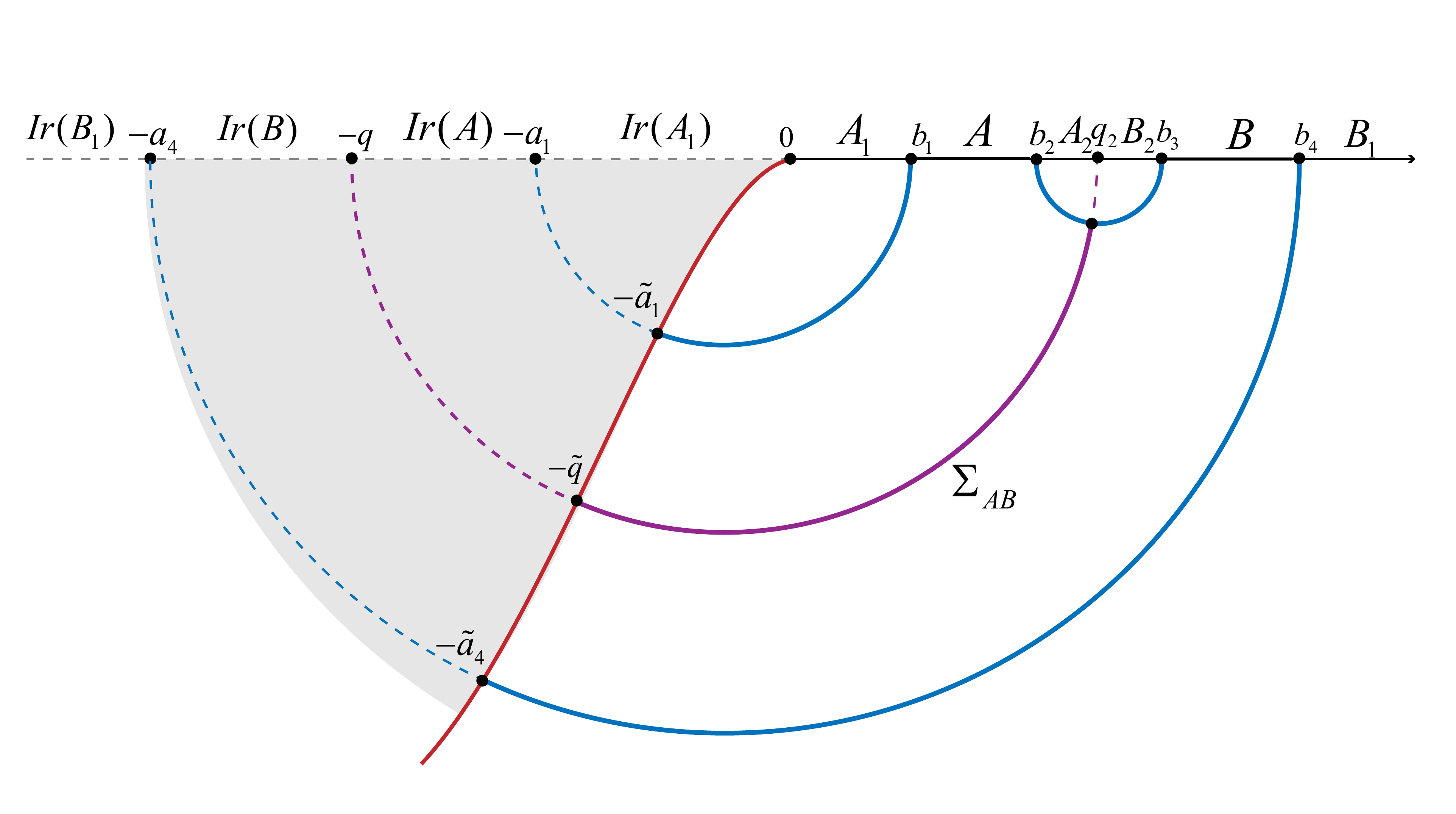}
    \caption{Phase D2b
    }
    \label{fig:jt-D2b}
\end{figure}

\begin{figure}
    \centering
    \includegraphics[width=0.8\textwidth]{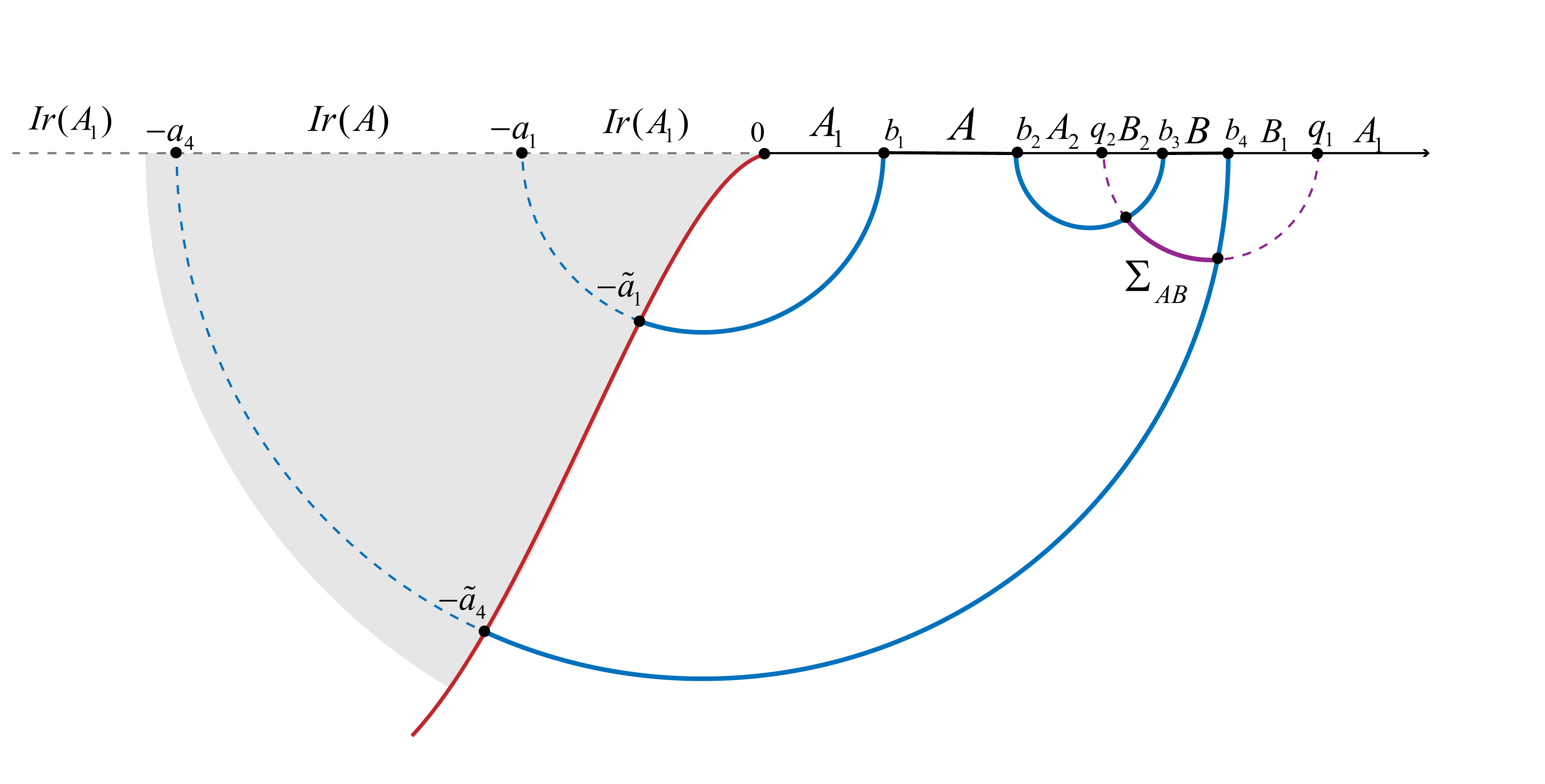}
    \caption{Phase D2c
    }
    \label{fig:jt-D2c}
\end{figure}

Now we consider the disjoint case where the island region of $AB$ in Weyl transformed CFT is given by $\text{Is}(AB)=[-a_4,-a_1]$ where $a_i\approx b_i+2x_0$ with $i=1,4$. 
And the three different assignments of the ownerless island are
\begin{enumerate}
    \item [1:] $\operatorname{Ir}(A)=\emptyset, \quad \operatorname{Ir}(B)=[-a_4,-a_1]$.
    \item [2:] $\operatorname{Ir}(A)=[-q,-a_1], \quad \operatorname{Ir}(B)=[-a_4,-q]$ with $a_1<q<a_4$.
    \item [3:] $\operatorname{Ir}(A)=[-a_4,-a_1], \quad \operatorname{Ir}(B)=\emptyset$.
\end{enumerate}
Akin to adjacent case, the above assignments lead to three distinct results for BPE, which match the areas of three saddles of the cross-section shown in Fig. \ref{fig:jt-D2a}, \ref{fig:jt-D2b} and \ref{fig:jt-D2c}, respectively.

\textbf{In Phase D2a}, the EWCS anchors on the RT($b_1$) and the RT surface of the interval $[b_2,b_3]$ and we should consider the assignment $\operatorname{Ir}(A)=\emptyset,\ \operatorname{Ir}(B)=[-a_4,-a_1]$.
Let us assume
\begin{equation}
   0<q_1<b_1,\quad \text{Ir}(A_1)=\emptyset,\quad
    \text{Ir}(B_1)=(-\infty,-a_4]\cup[-a_1,0].
\end{equation}

Now we solve the two balance conditions.
By the generalized ALC formula \eqref{galc2}, the four PEEs are given by
\begin{equation}
\begin{aligned}
 \mathcal{I}\left(A, B \operatorname{Ir}(B) B_1 \operatorname{Ir}\left(B_1\right) B_2\right) 
= & \frac{1}{2}\left[\tilde{S}_{A A_1}+\tilde{S}_{A_2}-\tilde{S}_{A_1}-\tilde{S}_{A_2}\right] \\
= & \frac{1}{2}\left[\tilde{S}_{\left[q_1, b_2\right]}+\tilde{S}_{\left[b_1, q_2\right]}-\tilde{S}_{\left[q_1, b_1\right]}-\tilde{S}_{\left[b_2, q_2\right]}\right] \\
= & \frac{c}{6} \log \frac{b_2-q_1}{q_2-b_2} \frac{q_2-b_1}{b_1-q_1},\\
 \mathcal{I}\left(B \operatorname{Ir}(B), A_1 A_2 A\right)
= & \frac{1}{2}\left[\tilde{S}_{B \operatorname{Ir}(B) B_1 \operatorname{Ir}\left(B_1\right)}+\tilde{S}_{B \operatorname{Ir}(B) B_2}-\tilde{S}_{B_2}-\tilde{S}_{B_1 \operatorname{Ir}\left(B_1\right)}\right] \\
= & \frac{c}{6} \log \frac{b_3-q_1}{b_3-q_2} \frac{q_2+a_1}{q_1+a_1},
\end{aligned}
\end{equation}
and
\begin{equation}
\begin{aligned}
\mathcal{I}\left(B_2, A_1 A_2 A\right)
= & \frac{1}{2}\left[\tilde{S}_{B \operatorname{Ir}(B) B_1 \operatorname{Ir}\left(B_1\right) B_2}+\tilde{S}_{B_2}-\tilde{S}_{B \operatorname{Ir}(B) B_1 \operatorname{Ir}\left(B_1\right)}\right] \\
= & \frac{1}{2}\left[\tilde{S}_{\left[q_1, q_2\right]}+\tilde{S}_{\left[q_2, b_3\right]}-\tilde{S}_{\left[q_1, b_3\right]}\right] \\
= & \frac{c}{6} \log \frac{\left(q_2-q_1\right)\left(b_3-q_2\right)}{\epsilon\left(b_3-q_1\right)}, \\
\mathcal{I}\left(A_2, B \operatorname{Ir}(B) B_1 \operatorname{Ir}\left(B_1\right) B_2\right)
= & \frac{1}{2}\left[\tilde{S}_{A A_1 A_2}+\tilde{S}_{A_2}-\tilde{S}_{A A_1}\right] \\
= & \frac{1}{2}\left[\tilde{S}_{\left[q_1, q_2\right]}+\tilde{S}_{\left[b_2, q_2\right]}-\tilde{S}_{\left[q_1, b_2\right]}\right] \\
= & \frac{c}{6} \log \frac{\left(q_2-q_1\right)\left(q_2-b_2\right)}{\left(b_2-q_1\right) \epsilon} .
\end{aligned}
\end{equation}
Solving the balanced conditions, we get the balanced points
\begin{equation}\label{d2a-bp-1}
    \begin{split}
        q_1&\approx \frac{b_1^2+b_2b_3-\sqrt{(b_2^2-b_1^2)(b_3^2-b_1^2)}}{b_2+b_3}
        -\left(
        \frac{2(b_1^2+b_2b_3-\sqrt{(b_2^2-b_1^2)(b_3^2-b_1^2)})}{(b_2+b_3)^2}\right.\\
        &\left.+\frac{-2b_1(b_1+b_2)(b_1+b_3)+\sqrt{(b_2^2-b_1^2)(b_3^2-b_1^2)}(2b_1+b_2+b_3)}{(b_2+b_3)(b_1+b_2)(b_1+b_3)}
        \right)x_0,
    \end{split}
\end{equation}
\begin{equation}\label{d2a-bp-2}
    \begin{split}
        q_2&\approx \frac{b_1^2+b_2b_3+\sqrt{(b_2^2-b_1^2)(b_3^2-b_1^2)}}{b_2+b_3}
        -\left(
        \frac{2(b_1^2+b_2b_3+\sqrt{(b_2^2-b_1^2)(b_3^2-b_1^2)})}{(b_2+b_3)^2}\right.\\
        &\left.+\frac{-2b_1(b_1+b_2)(b_1+b_3)-\sqrt{(b_2^2-b_1^2)(b_3^2-b_1^2)}(2b_1+b_2+b_3)}{(b_2+b_3)(b_1+b_2)(b_1+b_3)}
        \right)x_0,
    \end{split}
\end{equation}
which align with the two intersection points \eqref{tq1} and \eqref{tq2}.
Then the BPE is obtained as
\begin{equation}\label{bpe-d2a}
    \begin{split}
        \text{BPE}&=
        \frac{c}{6} \log \frac{b_2-q_1}{q_2-b_2} \frac{q_2-b_1}{b_1-q_1}\\
        &\approx\frac{c}{6} \log\left( \frac{b_2 b_3-b_1^2+\sqrt{\left(b_2^2-b_1^2\right)\left(b_3^2-b_1^2\right)}}{b_1\left(b_3-b_2\right)}\right)
-\frac{c}{6}\frac{\sqrt{\left(b_2^2-b_1^2\right)\left(b_3^2-b_1^2\right)}}{b_1(b_1+b_2)(b_1+b_3)}x_0,
    \end{split}
\end{equation}
which matches the area of EWCS \eqref{ew-d2a}.

\textbf{In Phase D2b}, the EWCS anchors on the brane and the RT surface of the interval $[b_2,b_3]$ and we should consider the assignment 
$\operatorname{Ir}(A)=[-q,-a_1],\ \operatorname{Ir}(B)=[-a_4,-q]$.
Assume that both $A_1$ and $B_1$ admit island contributions $\text{Ir}(A_1)$ and $\text{Ir}(B_1)$ and set
\begin{equation}
    A_1=[0,a_1],\quad 
    B_1=[a_4,+\infty),
\end{equation}
and accordingly
\begin{equation}
    \text{Ir}(A_1)=[-a_1,0],\quad
    \text{Ir}(B_1)=(-\infty,-a_4].
\end{equation}
By the generalized ALC formula \eqref{galc2}, the four PEEs are given by
\begin{equation}
\begin{aligned}
& \mathcal{I}\left(B \operatorname{Ir}(B), A \operatorname{Ir}(A) A_1 \operatorname{Ir}\left(A_1\right) A_2\right) \\
= & \frac{1}{2}\left[\tilde{S}_{B \operatorname{Ir}(B) B_1 \operatorname{Ir}\left(B_1\right)}+\tilde{S}_{B_2 B \operatorname{Ir}(B)}-\tilde{S}_{B_2}-\tilde{S}_{B_1 \operatorname{Ir}\left(B_1\right)}\right] \\
= & \frac{1}{2}\left(\tilde{S}_{[-\infty,-q] \cup\left[b_3,+\infty\right]}+\tilde{S}_{\left[-b_4,-q\right] \cup\left[q_2, b_4\right]}-\tilde{S}_{\left[q_2, b_3\right]}-\tilde{S}_{\left[-\infty,-b_4\right] \cup\left[b_4,+\infty\right]}\right) \\
= & \frac{c}{6} \log \frac{\left(b_3+q\right)\left(q+q_2\right)}{2q\left(b_3-q_2\right)}+\frac{c}{6} \rho_0+\frac{c}{6}\frac{x_0}{q},\\
& \mathcal{I}\left(A \operatorname{Ir}(A), B \operatorname{Ir}(B) B_1 \operatorname{Ir}\left(B_1\right) B_2\right) \\
= & \frac{1}{2}\left[\tilde{S}_{A \operatorname{Ir}(A) A_1 \operatorname{Ir}\left(A_1\right)}+\tilde{S}_{A_2 A \operatorname{Ir}(A)}-\tilde{S}_{A_1 \operatorname{Ir}\left(A_1\right)}-\tilde{S}_{A_2}\right] \\
= & \frac{1}{2}\left(\tilde{S}_{\left[-q, b_2\right]}+\tilde{S}_{\left[-q,-b_1\right] \cup\left[b_1, q_2\right]}-\tilde{S}_{\left[-b_1, b_1\right]}-\tilde{S}_{\left[b_2, q_2\right]}\right) \\
= & \frac{c}{6} \log \frac{\left(b_2+q\right)\left(q_2+q\right)}{2 q\left(q_2-b_2\right)}+\frac{c}{6} \rho_0+\frac{c}{6}\frac{x_0}{q},
\end{aligned}
\end{equation}
and 
\begin{equation}
\begin{aligned}
&\mathcal{I}\left(B_2, A \operatorname{Ir}(A) A_1 \operatorname{Ir}\left(A_1\right) A_2\right) \\
= & \frac{1}{2}\left[\tilde{S}_{B B_1 \operatorname{Ir}\left(B B_1\right) B_2}+\tilde{S}_{B_2}-\tilde{S}_{B B_1 \operatorname{Ir}\left(B B_1\right)}\right] \\
= & \frac{1}{2}\left(\tilde{S}_{[-\infty,-q] \cup\left[q_2, \infty\right]}+\tilde{S}_{\left[q_2, b_3\right]}-\tilde{S}_{[-\infty,-q] \cup\left[b_3,+\infty\right]}\right) \\
= & \frac{c}{6} \log \frac{\left(q_2+q\right)\left(b_3-q_2\right)}{\epsilon\left(b_3+q\right)},\\
&\mathcal{I}\left(A_2, B \operatorname{Ir}(B) B_1 \operatorname{Ir}\left(B_1\right) B_2\right) \\
= & \frac{1}{2}\left[\tilde{S}_{A \operatorname{Ir}(A) A_1 \operatorname{Ir}\left(A_1\right) A_2}+\tilde{S}_{A_2}-\tilde{S}_{A \operatorname{Ir}(A) A_1 \operatorname{Ir}\left(A_1\right)}\right] \\
= & \frac{1}{2}\left(\tilde{S}_{\left[-q, q_2\right]}+\tilde{S}_{\left[b_2, q_2\right]}-\tilde{S}_{\left[-q, b_2\right]}\right) \\
= & \frac{c}{6} \log \frac{\left(q_2+q\right)\left(q_2-b_2\right)}{\epsilon\left(b_2+q\right)} .
\end{aligned}
\end{equation}
We find that the two balance conditions are degenerate: 
\begin{equation}\label{bc-d2b}
\frac{b_3+q}{b_3-q_2}=\frac{b_2+q}{q_2-b_2}\quad 
\Rightarrow\quad q_2=\frac{2b_2b_3+(b_2+b_3)q}{b_2+b_3+2q}.
\end{equation}
Like the phase A2b, the BPE changes with $q$.
By varying $q$, we find the minimal BPE
\begin{equation}
    \begin{split}
     \text{BPE}(A:B)=  \frac{c}{6}\log\frac{\sqrt{b_3}+\sqrt{b_2}}{\sqrt{b_3}-\sqrt{b_2}}+\frac{c}{6} \rho_0+\frac{c}{6}\frac{x_0}{\sqrt{b_2b_3}},
    \end{split}
\end{equation}
which matches the area of EWCS \eqref{ew-d2b}.
The extremal point is given by
\begin{equation}\label{q-d2b}
    q\approx \sqrt{b_2b_3}+3(\sqrt{b_2}+\sqrt{b_3})^2\frac{x_0}{6\sqrt{b_2b_3}}, 
\end{equation}
Inserting \eqref{q-d2b} to \eqref{bc-d2b}, we get the corresponding balance point
\begin{equation}
    q_2\approx\sqrt{b_2b_3}+3(\sqrt{b_3}-\sqrt{b_2})^2\frac{x_0}{6\sqrt{b_2b_3}},
\end{equation}
which is exactly the intersection point \eqref{inter-d2b} by extending the EWCS to the asymptotic  boundary.

\textbf{In Phase D2c}, the EWCS anchors on the RT($b_4$) and the RT surface of the interval $[b_2,b_3]$ and we should consider the assignment $\operatorname{Ir}(B)=\emptyset,\ \operatorname{Ir}(A)=[-a_4,-a_1]$.
Following the similar procedure as in phase D2a, we get two balance points,
\begin{equation}
    \begin{split}
        q_1&\approx \frac{b_4^2+b_2b_3+\sqrt{(b_2^2-b_4^2)(b_3^2-b_4^2)}}{b_2+b_3}
        -\left(
        \frac{2(b_4^2+b_2b_3+\sqrt{(b_2^2-b_4^2)(b_3^2-b_4^2)})}{(b_2+b_3)^2}\right.\\
        &\left.+\frac{-2b_4(b_4+b_2)(b_4+b_3)-\sqrt{(b_2^2-b_4^2)(b_3^2-b_4^2)}(2b_4+b_2+b_3)}{(b_2+b_3)(b_4+b_2)(b_4+b_3)}
        \right)x_0,
    \end{split}
\end{equation}
\begin{equation}
    \begin{split}
       q_2&\approx \frac{b_4^2+b_2b_3-\sqrt{(b_2^2-b_4^2)(b_3^2-b_4^2)}}{b_2+b_3}
        -\left(
        \frac{2(b_4^2+b_2b_3-\sqrt{(b_2^2-b_4^2)(b_3^2-b_4^2)})}{(b_2+b_3)^2}\right.\\
        &\left.+\frac{-2b_4(b_4+b_2)(b_4+b_3)+\sqrt{(b_2^2-b_4^2)(b_3^2-b_4^2)}(2b_4+b_2+b_3)}{(b_2+b_3)(b_4+b_2)(b_4+b_3)}
        \right)x_0,
    \end{split}
\end{equation}
which are exactly the intersection points \eqref{D2c-tq1} and \eqref{D2c-tq2}.
The corresponding BPE is 
\begin{equation}
\begin{aligned}
\text{BPE}(A:B)
&=\frac{c}{6} \log\left( \frac{b_2 b_3-b_4^2+\sqrt{\left(b_2^2-b_4^2\right)\left(b_3^2-b_4^2\right)}}{b_4\left(b_3-b_2\right)}\right)-\frac{c}{6}\frac{\sqrt{\left(b_2^2-b_4^2\right)\left(b_3^2-b_4^2\right)}}{b_4(b_4+b_2)(b_4+b_3)}x_0,
\end{aligned}
\end{equation}
which coincides with the area of EWCS \eqref{ew-d2c}.

\subsection*{Minimizing BPE for phase D2}
Now let us compare the BPEs in phase D2.

The critical point between phase D2a and D2b is 
\begin{equation}
\begin{split}
F(b_1,b_2,b_3)=0,
\end{split}
\end{equation}
where 
\begin{equation}
  F(b_1,b_2,b_3)\equiv  \log\frac{b_3 b_2-b_1^2+\sqrt{\left(b_3^2-b_1^2\right)\left(b_2^{2}-b_1^2\right)}}{b_1\left(\sqrt{b_3}+\sqrt{b_2}\right)^2}
-\frac{x_0}{\sqrt{b_2b_3}}-\frac{x_0\sqrt{\left(b_2^2-b_1^2\right)\left(b_3^2-b_1^2\right)}}{b_1(b_1+b_2)(b_1+b_3)}
-\rho_0.
\end{equation}
When $F<0$, phase D2a dominates and gives the smaller BPE.
In fact, $F<0$ is also the condition that $A_1AA_2$ admits no island.
Now we compare the entanglement entropy for $A_1AA_2$ in no-island and island saddles, 
\begin{equation}\label{d2a-d2b-mbpe}
\begin{aligned}
&\frac{3}{c}(S_{\mathrm{non-island}}\left(A_1 A A_2\right)-S_{\mathrm{island}}\left(A_1 A A_2\right))  \\
=&\frac{1}{2}\log \frac{\left(q_2-q_1\right)^2}{4 q_1 q_2}-\frac{1}{2}\left(\frac{1}{q_2}+\frac{1}{q_1}\right)x_0-\rho_0 \\
=&\log \frac{\sqrt{\left(b_2^{ 2}-b_1^2\right)\left(b_3^2-b_1^2\right)}}{b_1\left(b_2+b_3\right)}-
\frac{(2b_1+b_2+b_3)(b_1^2+b_2b_3)x_0}{b_1(b_1+b_2)(b_1+b_3)(b_3+b_2)
}-\rho_0\\
\equiv& G(b_1,b_2,b_3).
\end{aligned}
\end{equation}
We observe that $G(b_1,b_2,b_3)-F(b_1,b_2,b_3)$ increases as $b_2\to b_3$, which indicates that
\begin{equation}\label{gf-ieq}
    G(b_1,b_2,b_3)-F(b_1,b_2,b_3)
    <G(b_1,b_3,b_3)-F(b_1,b_3,b_3)=0.
\end{equation}
Inserting \eqref{gf-ieq} to \eqref{d2a-d2b-mbpe}, we arrive at
\begin{equation}
\frac{3}{c}(S_{\mathrm{non-island}}\left(A_1 A A_2\right)-S_{\mathrm{island}}\left(A_1 A A_2\right))<F(b_1,b_2,b_3)<0.
\end{equation}
This is consistent with our assumption that $A_1 A A_2$ admits no island.

Likewise, for the BPEs between phase-D2b and phase-D2c, we conclude that phase-D2c gives the smaller BPE and $B_1BB_2$ admits no island when $F(b_4,b_2,b_3)<0$.

\subsection{BPE from minimizing the crossing PEE}
The balance point was shown to be the minimized point of the crossing PEE in vacuum CFT$_2$ \cite{Camargo:2022mme}, the covariant CFT$_2$ scenarios with or without gravitational anomaly \cite{Wen:2022jxr}  and the island phase with the unfluctuating brane \cite{Basu:2023wmv}. 
Moreover, the minimized crossing PEE for adjacent phase in CFT$_2$ is shown to be universal and exactly produce the Markov gap \cite{Camargo:2022mme}.
In fact, the minimization condition of the crossing PEE seems rather essential than the balanced requirement, as it may be even well-defined in more general scenarios like the discrete system. 

In this subsection, we would like to see if the minimized point is still the balanced point when the brane gets perturbed.
Without loss of generality, we will take phase A2a and phase D2a as examples.

For phase A2a, two crossing PEEs are 
\begin{equation}
\begin{aligned}
\mathcal{I}\left(A, B_1 \operatorname{Ir}\left(B_1\right)\right) & =\frac{1}{2}\left(\tilde{S}_{A A_1}+\tilde{S}_{A B \operatorname{Ir}(B)}-\tilde{S}_{A_1}-\tilde{S}_{B \operatorname{Ir}(B)}\right) \\
& =\frac{1}{2}\left(\tilde{S}_{\left[q_1, b_2\right]}+\tilde{S}_{\left[-a_4,-a_1\right] \cup\left[b_1, b_4\right]}-\tilde{S}_{\left[q_1, b_1\right]}-\tilde{S}_{\left[-a_4,-a_1\right] \cup\left[b_2, b_4\right]}\right) \\
& =\frac{c}{6} \log \left[\frac{(b_1+a_1)\left(b_2-q_1\right)}{\left(a_1+b_2\right)\left(b_1-q_1\right)}\right],\\
	\mathcal{I}(B\,\text{Ir}(B),A_1)&=\frac{1}{2}\left(\tilde{S}_{B\,\text{Ir}(B)\cup B_1\,\text{Ir}(B_1)}+\tilde S_{AB\text{Ir}(B)}-\tilde{S}_{A}-\tilde{S}_{B_1\text{Ir}(B_1)}\right)\\
	&=\frac{1}{2}\left(\tilde{S}_{(-\infty,q_1]\cup[b_2,\infty)}+\tilde{S}_{[-a_4,-a_1]\cup[b_1,b_4]}-\tilde{S}_{[b_1,b_2]}-\tilde{S}_{[-a_1,q_1]\cup (-\infty,-a_4]\cup[b_4,\infty)}\right)\\
	&=\frac{c}{6}\log\left[\frac{(b_1+a_1)(b_2-q_1)}{(b_2-b_1)(a_1+q_1)}\right],
\end{aligned}
\end{equation}
where $a_1=b_1+2x_0$.
Then the total crossing PEE is 
\begin{align}
\frac{1}{2}\Big[\mathcal{I}(A,B_1\text{Ir}(B_1))+\mathcal{I}(B\,\text{Ir}(B),A_1)\Big]&=\frac{c}{12}\log\left[\frac{(b_1+a_1)^2(b_2-q_1)^2}{(b_2-b_1)(a_1+q_1)\left(a_1+b_2\right)\left(b_1-q_1\right)}\right].
\end{align}
The minimal point for the total crossing PEE is
\begin{equation}
    q_1= \frac{b_1^2}{b_2}-\frac{(b_1-b_2)^2}{b_2^2}x_0,
\end{equation}
which is exactly the balanced point \eqref{bp-a2a}.
Furthermore, the minimized crossing PEE for this adjacent phase is 
\begin{equation}
\Big[\mathcal{I}(A,B_1\text{Ir}(B_1))+\mathcal{I}(B\,\text{Ir}(B),A_1)\Big]\Big|_{\rm minimal}
=\frac{c}{3}\log 2,
\end{equation}
which is universal and gives the lower bound of the Markov gap.
In island phase, the PEE for adjacent intervals does not recover the mutual information, and accordingly, the crossing PEE no longer exactly recovers the Markov gap \cite{Basu:2023wmv}.

For phase D2a, we have four crossing PEEs 
\begin{equation}
\begin{aligned}
\mathcal I(A,B_1\mathrm{Ir}(B_1))
&=\frac{1}{2}\left(\tilde{S}_{AA_1}+\tilde S_{AA_2B_2B\text{Ir}(B)}-\tilde{S}_{A_1}-\tilde{S}_{A_2B_2B\,\text{Ir}(B)}\right)\\
	&=\frac{1}{2}\left(\tilde{S}_{[q_1,b_2]}+\tilde{S}_{[-a_4,-a_1]\cup[b_1,b_4]}-\tilde{S}_{[q_1,b_1]}-\tilde{S}_{[-a_4,-a_1]\cup[b_2,b_4]}\right)\\
&=\frac{c}{6}\log\frac{(b_1+a_1)(b_2-q_1)}{(a_1+b_2)(b_1-q_1)},\\
\mathcal{I}(A,B_2)
&=\frac{1}{2}\left(\tilde{S}_{AA_2}+\tilde S_{B\text{Ir}(B)B_1\text{Ir}(B_1)A_1A}-\tilde{S}_{A_2}-\tilde{S}_{B\text{Ir}(B)B_1\text{Ir}(B_1)A_1}\right)\\
	&=\frac{1}{2}\left(\tilde{S}_{[b_1,q_2]}+\tilde{S}_{[b_2,b_3]}-\tilde{S}_{[b_2,q_2]}-\tilde{S}_{[b_1,b_3]}\right)\\
&=\frac{c}{6}\log\frac{(b_3-b_2)(q_2-b_1)}{(b_3-b_1)(q_2-b_2)},
\end{aligned}
\end{equation}
and
\begin{equation}
\begin{aligned}
\mathcal{I}(B\mathrm{Ir}(B),A_1)
&=\frac{1}{2}\left(\tilde{S}_{AA_2B_2B\text{Ir}(B)}+\tilde S_{B\text{Ir}(B)B_1\text{Ir}(B_1)}-\tilde{S}_{AA_2B_2}-\tilde{S}_{B_1\text{Ir}(B_1)}\right)\\
	&=\frac{1}{2}\left(\tilde{S}_{[-a_4,-a_1]\cup[b_1,b_4]}+\tilde{S}_{[q_1,b_3]}-\tilde{S}_{[b_1,b_3]}-\tilde{S}_{[-\infty,-a_4]\cup[-a_1,q_1]\cup[b_4,\infty]}\right)\\
&=\frac{c}{6}\log\frac{(a_1+b_1)(b_3-q_1)}{(a_1+q_1)(b_3-b_1)},\\
\mathcal{I}(B\mathrm{Ir}(B),A_2)
&=\frac{1}{2}\left(\tilde{S}_{B_2B\text{Ir}(B)}+\tilde S_{B\text{Ir}(B)B_1\text{Ir}(B_1)A_1A}-\tilde{S}_{B_2}-\tilde{S}_{B_1\text{Ir}(B_1)A_1A}\right)\\
	&=\frac{1}{2}\left(\tilde{S}_{[-a_4,-a_1]\cup[q_2,b_4]}+\tilde{S}_{[b_2,b_3]}-\tilde{S}_{[q_2,b_3]}-\tilde{S}_{[-a_4,-a_1]\cup[b_2,b_4]}\right)\\
&=\frac{c}{6}\log\frac{(q_2+a_1)(b_3-b_2)}{(a_1+b_2)(b_3-q_2)},
\end{aligned}
\end{equation}
Then the total crossing PEE is 
\begin{equation}
\begin{aligned}
&\mathcal I(A,B_1\mathrm{Ir}(B_1))+
\mathcal{I}(A,B_2)
+\mathcal{I}(B\mathrm{Ir}(B),A_1)+
\mathcal{I}(B\mathrm{Ir}(B),A_2)\\
=&\frac{c}{6}\log\left[\frac{(a_1+b_1)^2(b_3-b_2)^2(q_2-b_1)(q_2+a_1)(b_2-q_1)(b_3-q_1)}{(b_2+a_1)^2(b_3-b_1)^2(b_1-q_1)(a_1+q_1)(q_2-b_2)(b_3-q_2)}\right].
\end{aligned}
\end{equation}
The minimal points are given by 
\begin{equation}
    \begin{split}
        q_1&\approx \frac{b_1^2+b_2b_3-\sqrt{(b_2^2-b_1^2)(b_3^2-b_1^2)}}{b_2+b_3}
        -\left(
        \frac{2(b_1^2+b_2b_3-\sqrt{(b_2^2-b_1^2)(b_3^2-b_1^2)})}{(b_2+b_3)^2}\right.\\
        &\left.+\frac{-2b_1(b_1+b_2)(b_1+b_3)+\sqrt{(b_2^2-b_1^2)(b_3^2-b_1^2)}(2b_1+b_2+b_3)}{(b_2+b_3)(b_1+b_2)(b_1+b_3)}
        \right)x_0,
    \end{split}
\end{equation}
\begin{equation}
    \begin{split}
        q_2&\approx \frac{b_1^2+b_2b_3+\sqrt{(b_2^2-b_1^2)(b_3^2-b_1^2)}}{b_2+b_3}
        -\left(
        \frac{2(b_1^2+b_2b_3+\sqrt{(b_2^2-b_1^2)(b_3^2-b_1^2)})}{(b_2+b_3)^2}\right.\\
        &\left.+\frac{-2b_1(b_1+b_2)(b_1+b_3)-\sqrt{(b_2^2-b_1^2)(b_3^2-b_1^2)}(2b_1+b_2+b_3)}{(b_2+b_3)(b_1+b_2)(b_1+b_3)}
        \right)x_0,
    \end{split}
\end{equation}
which is exactly the balanced points \eqref{d2a-bp-1} and \eqref{d2a-bp-2}.

\section{Discussion}\label{sec:dis}
In this paper we simulate the AdS/BCFT correspondence with a fluctuating KR brane via a holographic Weyl transformed CFT$_2$, where the Weyl transformation is adapted to induce a cutoff brane that overlaps with the KR brane. We calculated the entanglement entropy in the Weyl transformed CFT$_2$ and the corresponding RT surface and find it coincides with the RT surface in AdS/BCFT. Furthermore, we calculated the BPE in the Weyl transformed CFT by taking into account of the island contributions, and find its correspondence to the EWCS in the holographic picture. Compared with the calculation in \cite{Basu:2023wmv}, the fluctuation of the brane changes both of the BPE and the EWCS. Provided different assignments for the ownerless island region, we solve the balance requirements and find that, the corresponding BPE matches the area of different saddle points of the EWCS. According to the minimal requirement for the BPE, we ought to select the assignment that yields the minimal BPE, which corresponds to the area of the EWCS.

Although the simulation is quite successful in reproducing the RT surfaces anchored on the brane and the EWCS in the entanglement wedges via field theory calculations in the Weyl transformed CFT, there are important differences between the holographic Weyl  transformed CFT and the AdS/BCFT set-up.
\begin{itemize}
\item Firstly, the Weyl transformed CFT is always settled at the asymptotic boundary, while the $2d$ effective theory in AdS/BCFT locates on the KR brane and the bath CFT.

\item Secondly, the spacetime behind the cutoff brane is not empty. The cutoff brane is just the boundary of the region where the RT curves are cut off. We stress that, when calculating a two-point function of the twist operators, the corresponding RT curve should always be cut off at some point on the corresponding cutoff sphere, which in general is not the point on the cutoff brane. For an arbitrary two-point function, the corresponding RT curve are cut off at some point behind the cutoff brane, and only the RT curves representing the optimized two-point functions are cut off at the cutoff brane.  
\end{itemize}

Compared with the AdS/BCFT (or JT gravity coupled to a CFT bath) set-up, the holographic Weyl transformed CFT$_2$ has many advantages:
\begin{itemize}
\item The two-point functions of twist operators without optimization are subtle in the AdS/BCFT set-up, while they are well defined in the Weyl transformed CFT. It was pointed out that, the non-optimized two-point functions should not be understood as an entanglement entropy due to the self-encoding property, rather it is proposed to be a PEE between the interval bounded by the two points and its complement \cite{Basu:2022crn}.

\item In the AdS/BCFT set-up, since the JT gravity lives on the brane, the insertion point of the twist operator in the JT gravity and the point where the RT surface anchors on the brane should coincide. 
This is inconsistent with our observation that $a_{\rm brane}\neq a_{\rm bdy}$, which happens in both of the AdS/BCFT and the Weyl transformed CFT set-ups. Nevertheless, the inequality is well understood in the Weyl transformed CFT set-up, as the Weyl transformed CFT locates at the asymptotic boundary, rather than the brane. 

\item The Weyl factor 1 \eqref{Omega1} is a result of the conjecture that the $2d$ effective theory in AdS/BCFT lives on the brane. Nevertheless, we showed that the island formula \eqref{island_EE} with the Weyl factor  \eqref{Omega1} gives us the entanglement entropy \eqref{Omega1Sa}, which deviates from the result \eqref{doublehsa} in doubly holography. While the island formula with the Weyl factor 2 \eqref{Omega2} reproduces \eqref{doublehsa}.
\end{itemize}

In \cite{Deng:2020ent}, the authors calculated the entanglement entropy for intervals in the bath region by the island formula \eqref{island_EE} with the Weyl factor 1 \eqref{Omega1}, and tried to match it to the result of a bulk formula, which they call the defect extremal surface (DES) formula. Nevertheless in the set-up of \cite{Deng:2020ent}, it is necessary to put specific defect theory on the brane by hand, and the coupling between the defect theory and the CFT on the bath region to form a $2d$ effective theory is subtle. It will be interesting to decompose the DES set-up using the Weyl transformed CFT set-up.

The particular Weyl transformations we have used are derived from the requirement that, the cutoff brane should coincide with the KR brane in the AdS/BCFT setup. Note that, in the AdS/BCFT set-up the location of the KR brane is determined by the boundary conditions and tension on the brane and the Einstein equation of the gravity. It will be very important to explore whether there is any intrinsic reason in the Weyl transformed CFT that reproduces the particular Weyl transformation.

In summary, our discussion on the configuration with a fluctuating brane indicates that, the Weyl transformed CFT gives us a more consistent interpretation for the island formulas of entanglement entropy and BPE, which further indicates that several features of the original AdS/BCFT setup should be reconsidered. Furthermore, the differences between the Weyl transformed CFT and the AdS/BCFT setup indicates that the $2d$ effective field theory should locate on the asymptotic boundary rather than the KR brane, which resolves the puzzle of $a_{\rm bdy}\neq a_{\rm brane}$ in the AdS/BCFT setup. Exploring the differences between these two setups and testing their self-consistency from other perspectives will be very interesting future directions.

\section*{Acknowledgment}
We would like to thank Hao Geng, Ziming Ji, Zhenbin Yang and Yang Zhou for helpful discussions. J. Lin is supported by the National Natural Science Foundation of China under Grant No.12247117, No.12247103 and No.12047502.
Y. Lu receives support from the China Postdoctoral Science Foundation under Grant No.2022TQ0140, the National Natural Science Foundation of
China under Grant No.12247161, and the NSFC Research Fund for International Scientists (Grant No. 12250410250).

\appendix
\section{Cutoff brane of Weyl transformed CFT}\label{app:cut}
\begin{figure}
    \centering
    \includegraphics[width=0.8\textwidth]{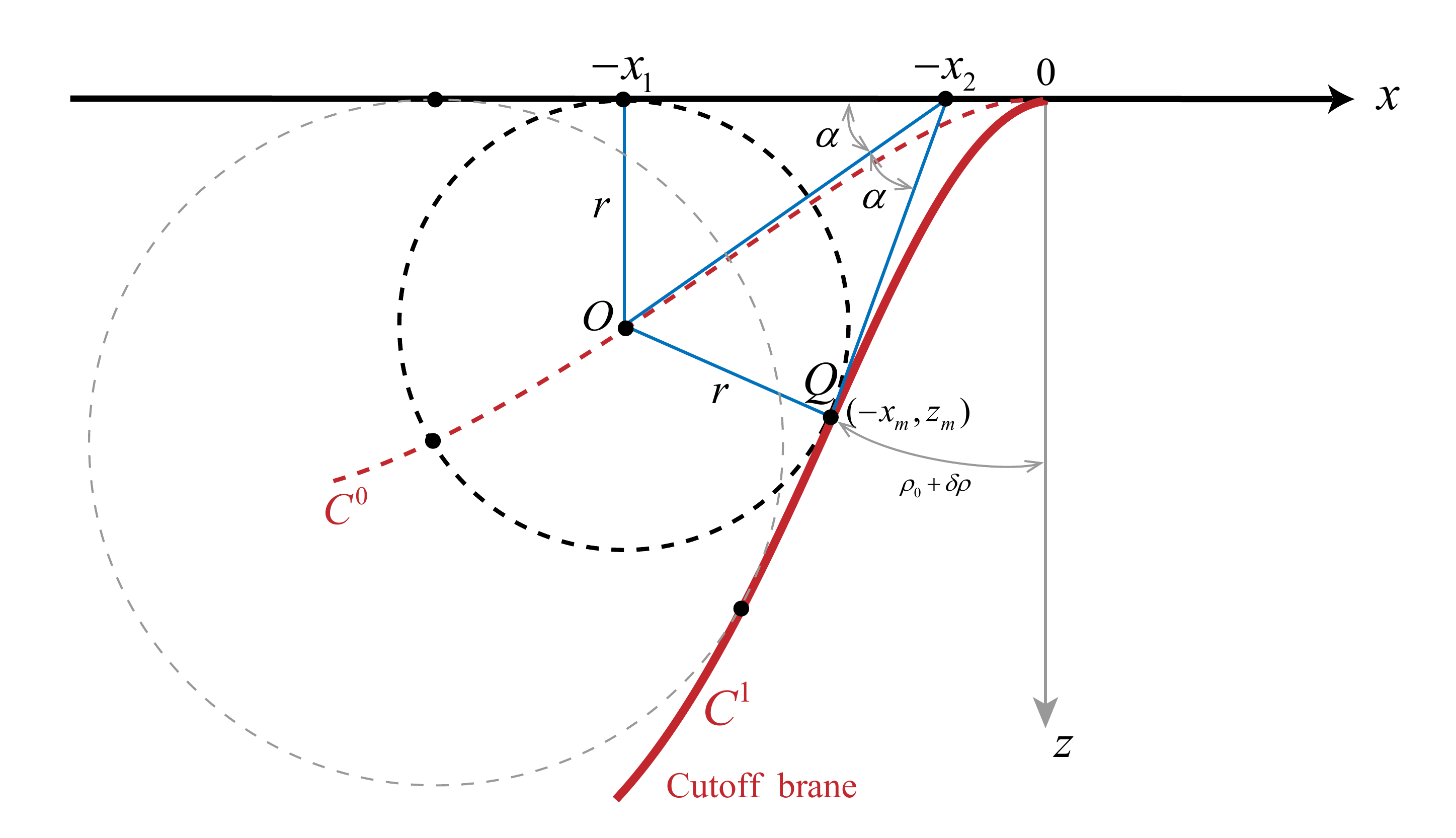}
    \caption{Cutoff brane (the red curve) in Weyl transformed CFT is the comment tangent line of the set of cutoff spheres.
    }
    \label{fig:cutoff-brane}
\end{figure}

We first solve the cutoff sphere associated with $x=-x_1$, which is defined as a set of points with the geodesic distance $|\varphi(-x_1)|$ from the boundary  $x=-x_1$, where
\begin{equation}
\varphi(x)= \begin{cases}0 & \text { if } \quad x>0 \\ -\log \left(\frac{2|x|}{\epsilon}\right)+\kappa_1+\frac{\kappa_2}{|x|}  & \ \text {if } \quad x<0\end{cases}
\end{equation}
The Poincaré AdS$_3$ metric is
\begin{equation}
\mathrm{d} s^2=\frac{-\mathrm{d} t^2+\mathrm{d} x^2+\mathrm{d} z^2}{z^2}.
\end{equation}
Using the geodesic length between two spacelike-separated points \cite{Wen:2018mev}
\begin{equation}
\begin{aligned}
& L_{\text {AdS }}\left(U_1, V_1, \rho_1, U_2, V_2, \rho_2\right) \\
& \quad=\frac{1}{2} \log \left[\frac{\rho_2\left(\rho_2+X\right)+\rho_1\left(\rho_2 Y\left(2 \rho_2+X\right)+X\right)+\left(\rho_1+\rho_2 \rho_1 Y\right)^2}{2 \rho_1 \rho_2}\right],
\end{aligned}
\end{equation}
where
\begin{equation}
U=\frac{x+t}{2}, \quad V=\frac{x-t}{2}, \quad \rho=\frac{2}{z^2},
\end{equation}
are light-cone coordinate and 
\begin{equation}
\begin{aligned}
& Y=2\left(U_1-U_2\right)\left(V_1-V_2\right), \quad
X=\sqrt{\rho_1^2+2 \rho_2 \rho_1\left(\rho_1 Y-1\right)+\left(\rho_2+\rho_1 \rho_2 Y\right)^2},
\end{aligned}
\end{equation}
we get the cutoff sphere associated with $x=-x_1$ on  a static time slice $t=0$
\begin{equation}
\left(x-x_1\right)^2+\left(z-x_1 e^{-\kappa_1-\frac{\kappa_2}{|x_1|}}\right)^2=x_1^2 e^{-2\kappa_1-2\frac{\kappa_2}{x_1}},
\end{equation}
which is just the circle at $(-x_1,x_1e^{-\kappa_1-\frac{\kappa_2}{x_1}})$ with radius $r=e^{-\kappa_1-\frac{\kappa_2}{x_1}}$.

Now we seek for the common tangent line $C^1$ of the set of cutoff spheres (see the illustration in Fig.\ref{fig:cutoff-brane}). 
This tangent line $C^1$ is what we call cutoff surface for Weyl transformed CFT.
The curve $C^0$, which is made up of the set of the center points of all the cutoff spheres, is given by
\begin{equation}
    C^0:\quad z=e^{-\kappa_1-\frac{\kappa_2}{|x|}}.
\end{equation}
Then the tangent line of $C^0$ at $O=(-x_1,r)$  and the tangent line of $C^1$ at $Q=(-x_m,z_m)$ will both intersect with the AdS boundary at 
\begin{equation}\label{tangentbrane}
    x=-x_2=-\frac{\kappa_2 x_1}{\kappa_2+x_1},
\end{equation}
with the intersection angle $\alpha$ and $2\alpha$, respectively and 
\begin{equation}
    \tan \alpha= e^{-(\kappa_1+\frac{\kappa_2}{x_1})}\left(\frac{\kappa_2}{x_1}+1\right).
\end{equation}
Then the tangent point $Q=(-x_m,z_m)$ is determined by the following geometric relations
\begin{equation}\label{geo}
    \frac{z_m}{x_m-x_2}=\tan 2\alpha,\quad
    \frac{x_m-x_2}{x_1-x_2}=\cos 2\alpha.
\end{equation}
Solving \eqref{geo}, we arrive at the parametric equation of the cutoff surface   
\begin{equation}\label{c1}
\begin{split}
    -x_m& \approx -x_1\tanh{\kappa_1}-(1-\tanh{\kappa_1})\kappa_2+\mathcal{O}(\kappa_2^2),\\
    z_m& \approx
    x_1\cosh^{-1}{\kappa_1}-\kappa_2\cosh^{-1}{\kappa_1}+\mathcal{O}(\kappa_2^2),
    \end{split}
\end{equation}
which is parameterized by $x_1$.

Now we show that the location of the cutoff brane \eqref{c1} coincide with the JT brane.
The location of JT brane is given by 
\begin{equation}\label{jt-locate}
    \begin{split}
        -x_J&=-|y|\tanh\left[ \rho_0\left(1+\frac{\bar\phi_r}{|y|}\right)\right]
        \approx 
        -|y|\tanh  \rho_0-
        \frac{\rho_0\bar\phi_r}{\ell}\cosh^{-2}\left( \rho_0\right)+\mathcal{O}(\bar\phi_r^2)\\
        z_J&=|y|\cosh^{-1}\left[ \rho_0\left(1+\frac{\bar\phi_r}{|y|}\right)\right]
        \approx
        |y| \cosh^{-1}  \rho_0-
        \frac{\rho_0\bar\phi_r}{\ell}\cosh^{-1}\left( \rho_0\right)\tanh\left( \rho_0\right)+\mathcal{O}(\bar\phi_r^2),
    \end{split}
\end{equation}
which is parameterized by $|y|=\sqrt{x_J^2+z_J^2}$, which is the Euclidean distance between a point $(-x_J,z_J)$ on the brane and the origin point.
To compare JT brane \eqref{jt-locate} with the cutoff brane \eqref{c1}, let us reparametrize the cutoff brane in terms of $|y|$. 
Using 
\begin{equation}
    \begin{split}
        |y|=\sqrt{x_m^2+z_m^2}
        \approx 
        x_1 +\kappa_2(\tanh\kappa_1-1)
        +\mathcal{O}(\kappa_2^2),
    \end{split}
\end{equation}
we get the reparametrized cutoff brane 
\begin{equation}\label{c11}
\begin{split}
    -x_m& \approx -|y|\tanh{\kappa_1}-\kappa_2\cosh^{-2}\kappa_1+\mathcal{O}(\kappa_2^2),\\
    z_m& \approx
    |y|\cosh^{-1}{\kappa_1}-\kappa_2\tanh\kappa_1\cosh^{-1}\kappa_1+\mathcal{O}(\kappa_2^2).
    \end{split}
\end{equation}
Identifying 
\begin{equation}
    \kappa_1= \rho_0,\quad
    \kappa_2=\frac{\rho_0\bar\phi_r}{\ell},
\end{equation}
then the cutoff brane coincides with the JT brane.

\section{A trick to calculate EWCS for phase D2b}\label{B}
\begin{figure}
    \centering
    \includegraphics[width=0.6\textwidth]{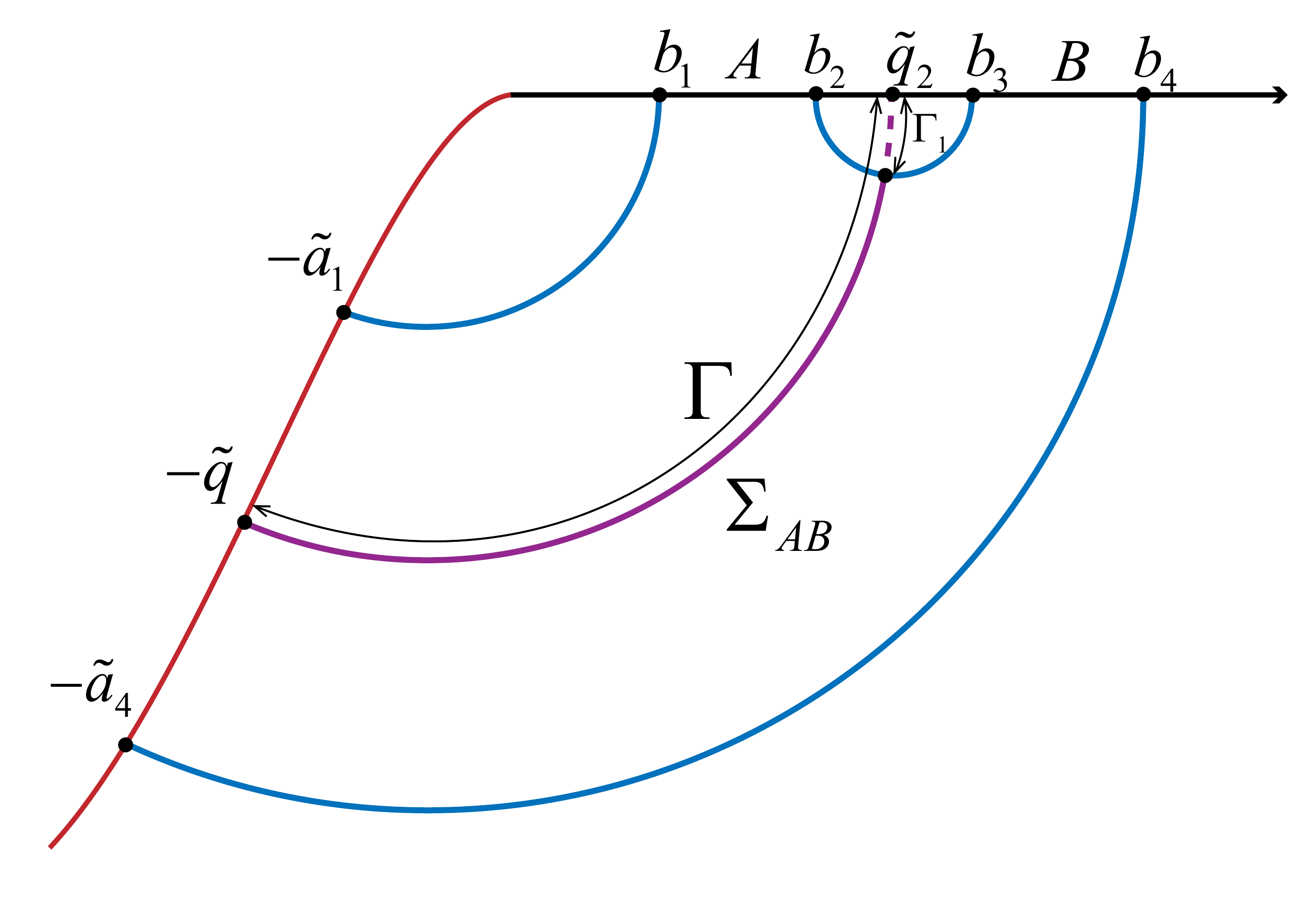}
    \caption{The area of EWCS can be obtained by the area of $\Gamma$ minus the area of $\Gamma_1$, where
$\Gamma$ is the extended EWCS and $\Gamma_1$ the extended part of $\Gamma$.
    }
    \label{fig:trick}
\end{figure}
Let us extend EWCS for phase D2b to the point $(\tilde q_2,0)$ on the AdS boundary (see Fig.\ref{fig:trick}). 
Then this extended EWCS $\Gamma$ is the RT surface connecting AdS boundary point $x=\tilde q_2$  and the point $y=\tilde q$ 
 on the brane  and its area is 
\begin{equation}
    \begin{split}
      \frac{\text{Area}(\Gamma)}{4G_N}
      =& \frac{c}{6}\log\frac{\tilde q_2^2+2 \tilde q_2 \tilde q \tanh \rho_0+\tilde q^2}{\tilde q_2 \epsilon\cosh^{-1} \rho_0}+\frac{c x_0}{6\tilde q},
    \end{split}
\end{equation}
where 
\begin{equation}
    \tilde q\approx \tilde q_2+x_0\left(1+\tanh \rho_0\right).
\end{equation}
Note that the EWCS is minimal curve and thus perpendicular to the RT surface associated with the interval $[b_2,b_3]$. 
Then, the area of the segment $\Gamma_1$ on $\Gamma$ that  starts from $x=\tilde q_2$ and ends at the RT surface for $[b_2,b_3]$ is given by
\begin{equation}
    \begin{split}
      \frac{\text{Area}(\Gamma_1)}{4G_N}
      =& \frac{c}{6}\log\frac{2(\tilde q_2-b_2)(b_3-\tilde q_2)}{(b_3-b_2)\epsilon}.
    \end{split}
\end{equation}
Finally, the area of EWCS is calculated by the following extremization procedure 
\begin{equation}
     \frac{\text{Area}(\text{EWCS})}{4G_N}=\text{Ext}_{\tilde q_2}\left(  \frac{\text{Area}(\Gamma)}{4G_N}- \frac{\text{Area}(\Gamma_1)}{4G_N}\right),     
\end{equation}
with the extremal point is 
\begin{equation}
     \tilde q_2=\sqrt{b_2b_3}+3(\sqrt{b_3}-\sqrt{b_2})^2\frac{x_0}{6\sqrt{b_2b_3}}.
     \end{equation}
Then the area of EWCS is
\begin{equation}
    \begin{split}
      \frac{\text{Area}(\text{EWCS})}{4G_N}=  \frac{c}{6}\log\frac{\sqrt{b_3}+\sqrt{b_2}}{\sqrt{b_3}-\sqrt{b_2}}+\frac{c}{6} \rho_0+
   \frac{c}{6}\frac{x_0}{\sqrt{b_2b_3}}.
    \end{split}
\end{equation}

\bibliography{ref.bib}
\bibliographystyle{SciPost_bibstyle}

\end{document}